\documentclass[10pt, twocolumn]{IEEEtran}
\usepackage{latexsym, graphicx}
\usepackage{amsmath,cases}
\usepackage{algorithm,algpseudocode,float}
\usepackage{epsfig}
\usepackage{amsfonts, amssymb, amsthm}
\usepackage{booktabs}
\usepackage{float} 
\usepackage{fancyhdr}
\usepackage{subfigure}  
\usepackage{graphicx}
\usepackage{color}
\usepackage{amssymb}
\usepackage{epstopdf}
\usepackage{amsfonts}
\usepackage{amsmath}
\usepackage{subfigure}
\usepackage{balance}
\usepackage{tabularx}
\usepackage{bm}
\usepackage{multirow}
\usepackage{cite}
\usepackage{url}
\usepackage{float}
\usepackage{color}

\usepackage{cite}
\usepackage{amsmath,amssymb,amsfonts}
\usepackage{verbatim}
\usepackage{graphicx}
\usepackage{textcomp}
\usepackage{amsmath,bm}
\usepackage{mathrsfs}
\usepackage{bm}
\def\BibTeX{{\rm B\kern-.05em{\sc i\kern-.025em b}\kern-.08em
		T\kern-.1667em\lower.7ex\hbox{E}\kern-.125emX}}

\usepackage{mathtools}
\newtagform{AppendixA}{(A}{)}
\newtagform{AppendixB}{(B}{)}
\newtagform{AppendixC}{(C}{)}

\makeatletter
\newenvironment{breakablealgorithm}
{
	\begin{center}
		\refstepcounter{algorithm}
		\hrule height.8pt depth0pt \kern2pt
		\renewcommand{\caption}[2][\relax]{
			{\raggedright\textbf{\ALG@name~\thealgorithm} ##2\par}%
			\ifx\relax##1\relax 
			\addcontentsline{loa}{algorithm}{\protect\numberline{\thealgorithm}##2}%
			\else 
			\addcontentsline{loa}{algorithm}{\protect\numberline{\thealgorithm}##1}%
			\fi
			\kern2pt\hrule\kern2pt
		}
	}{
		\kern2pt\hrule\relax
	\end{center}
}
\makeatother

\begin{document}
%
%

\title{Probabilistic Charging Power Forecast of EVCS: Reinforcement Learning Assisted Deep Learning Approach}

\author{Yuanzheng Li, {\it Member IEEE}, Shangyang He, Yang Li, {\it Senior Member IEEE}, \\ Leijiao Ge, Suhua Lou, 
and Zhigang Zeng, {\it Fellow IEEE} 

\thanks{This work was supported in part by the National Natural Science Foundation of China under Grant 62073148, in part by Natural Science Foundation of Jilin Province, China under Grant No. YDZJ202101ZYTS149, and in part by the 111 Project on Computational Intelligence and Intelligent Control under Grant B18024. (Corresponding author: Yang Li)} 		
	
\thanks{Y. Z. Li and Z. G. Zeng are with School of Artificial Intelligence and Automation, Key Laboratory on Image Information Processing and Intelligent Control of Ministry of Education, Huazhong University of Science and Technology,
Wuhan 430074, and also with China-Belt and Road Joint Laboratory on Measurement and Control Technology, Wuhan, China, 430074 (Email: Yuanzheng\_Li@hust.edu.cn, zgzeng@hust.edu.cn).}

\thanks{S. Y. He is with China-EU Institute for Clean and Renewable Energy, Huazhong University of Science and Technology, Wuhan 430074, China (Email: heshangyang10@hust.edu.cn).}

\thanks{Y. Li is with School of Electrical Engineering, Northeast Electric Power University, Jilin 132012, China (Email:liyang@neepu.edu.cn).}

\thanks{L. J. Ge is with the Key Laboratory of Smart Grid of Ministry of Education, Tianjin University,
Tianjin, 300072 China (Email:legendglj99@tju.edu.cn).}

\thanks{S. H. Lou is with School of Electrical and Electronic Engineering, Huazhong University of Science and Technology, Wuhan 430074, China (Email: shlou@mail.hust.edu.cn).}

}

\maketitle

\begin{abstract}
	The electric vehicle (EV) and electric vehicle charging station (EVCS) have been widely deployed with the development of large-scale transportation electrifications. However, since charging behaviors of EVs show large uncertainties, the forecasting of EVCS charging power is non-trivial. This paper tackles this issue by proposing a reinforcement learning assisted deep learning framework for the probabilistic EVCS charging power forecasting to capture its uncertainties. Since the EVCS charging power data are not standard time-series data like electricity load, they are first converted to the time-series format. {{On this basis, one of the most popular deep learning models, the long short-term memory (LSTM) is used and trained to obtain the point forecast of EVCS charging power.}} To further capture the forecast uncertainty, a Markov decision process (MDP) is employed to model the change of LSTM cell states, which is solved by our proposed adaptive exploration proximal policy optimization (AePPO) algorithm based on reinforcement learning. Finally, experiments are carried out on the real EVCSs charging data from Caltech, and Jet Propulsion Laboratory, USA, respectively. The results and comparative analysis verify the effectiveness and outperformance of our proposed framework.
\end{abstract}

\IEEEpeerreviewmaketitle

\section{Introduction}
\IEEEPARstart{A}{s} one of the most essential actions, the carbon emission reduction has been adopted to address the problem of climate change. Therefore, renewable energy sources, such as wind and solar energy, are widely deployed to substitute fossil fuels. Besides, the electrification of transportation systems is another effective approach to reduce carbon emissions \cite{abs_TIV1}.
Due to the development of manufacture and related key technologies, the reliability and efficiency of electric vehicles (EVs) are significantly improved. Therefore, various countries aim to achieve a high EV penetration in the near future. It is predicted that the number of EVs will increase continually and may approximately reach 35 million at the end of 2022 throughout the world \cite{liu_review_2015}.

{{However, as one of the most important factors to expand the EV adoption, EV charging has brought great challenges to the secure and economic operations of the EV charging station (EVCS). Because the charging behavior of EV is with the nature of intrinsic randomness, the charging power of EVCS is uncertain \cite{shi_model_2019}. Consequently, it is considered as a volatile and uncertain electricity load. Fluctuations of such load may threaten the operational security of EVCS and corresponding power systems \cite{chaudhari_agent-based_2019}. To prevent these issues, it is crucial to well predict the EVCS charging power.}}

{{Current works about EVCS or EV charging power forecasting can be divided into two main categories, namely, the model-based and data-driven approaches. Different from the wind power and residential load power forecasting that are mainly influenced by the natural environment and the living habits of residents, the charging behavior of the EV users is the focus on the first category.}}
Ref. \cite{cheng_charging_2020} has employed the trip chain to establish a spatial-temporal behavior model of EVs, which is beneficial for forecasting their charging power. Considering the impacts of traffic on the charging behavior, Ref. \cite{chen_spatial-temporal_2020} has provided a reliable approach for EVCS charging power forecasting. Based on the case in Shenzhen, China, the charging behaviors of EVs are considered for systematic forecasting of EV charging power \cite{ZHENG2020102084}.

Thanks to the advent of the cloud services and the internet of things, massive charging process data, such as the charging time and energy of EVs, can be collected. Thus, data-driven algorithms could be used for the forecast of EVCS charging power \cite{feng_review_2020}. {{Some classical data-driven forecasting algorithms are widely employed.}}
For instance, \cite{louie_time-series_2017} has applied an autoregressive integrated moving average (ARIMA) model to forecast the charging power of massive EVCSs, which were distributed in Washington State and San Diego. Ref. \cite{zhang_short-term_2018} has combined the least squares support vector machine algorithm and fuzzy clustering for the EVCS charging power forecasting task. Based on the historical data of Nebraska, USA, the effectiveness of the extreme gradient boosting on the charging power forecasting has been validated in \cite{almaghrebi_data-driven_2020}.

The above algorithms rely on the structured input data with a specific human-defined feature, which requires cumbersome engineering-based efforts \cite{zhu_electric_2019}. Hence, the deep learning method \cite{ZLLi} becomes more popular in this field. For instance, Ref. \cite{zhu_electric_2019} has reviewed numerous deep learning methods for EV charging load forecasting, and concluded that the long short-term memory (LSTM) model could reduce by about 30\% forecasting error compared with the conventional artificial neural networks (ANN) model. {{Considering the temporal dependency of the data, Ref. \cite{Editor_paper} has implemented the LSTM to capture the peak of charging power and achieve an effective result.}} Ref. \cite{xue_research_2021} has introduced the LSTM network to build a hybrid model for the forecasting of EVCS charging power and the experimental results demonstrated its effectiveness. Besides, compared with ARIMA and ANN, Ref. \cite{kim_forecasting_2021} has also illustrated the advantage of LSTM with abundant experiments.

Note that the above algorithms have been used to conduct the point forecasting, which only provides an expected charging power of EVCS in the future. Therefore, the probabilistic forecasting is introduced. It could forecast the probabilistic distribution of the future charging power, and thus provide more information, i.e., the expected value and the forecast uncertainties 
\cite{buzna_ensemble_2021}\cite{zhang_deep-learning-based_2021}. {{Consequently, the probabilistic forecasting of EVCS charging power is conducted in this paper. To the best of our knowledge, there exist only a few works about this topic.}} For instance, Ref. \cite{buzna_ensemble_2021} has applied four quantiles regression algorithms to the probabilistic EV load forecasting. In addition, the deep learning method is used in Ref. \cite{zhang_deep-learning-based_2021} to capture uncertainties of the probabilistic forecasting.

However, the above quantiles regression and deep learning-based methods have disadvantages. {{Regarding the first method, several steps are needed to be taken under different quantiles to obtain the forecast distribution, which would bring about much computational burden \cite{buzna_ensemble_2021}.}} For deep learning, the forecast uncertainty is mainly caused by the model training and input data \cite{zhang_deep-learning-based_2021}\cite{zhu_deep_2017}. Indeed, the stochastic parameters are normally introduced in the deep learning model, and the uncertainty is estimated through statistical indicators of its outputs. Therefore, the uncertainty of probabilistic forecast is usually obtained by using adequate samples, which needs the repeated running of the deep learning model and the usage of input data. In other words, traditional approaches may lead to a huge computation complexity.

Consequently, we propose an approach to solve the above issue. That is, a reinforcement learning assisted deep learning forecast approach is adopted to directly obtain the forecast uncertainty. This approach is designed to calculate the uncertainty once only, instead of repeated running. In detail, the LSTM is used to obtain the point prediction results of EVCS charging power, which is deemed as the expected value of the forecasting probabilistic distribution. Furthermore, note that the cell state of LSTM is one of the core inherent parameters, which could represent the important information of the model and the input data, simultaneously. Therefore, the forecast uncertainty can be obtained from the cell state (this issue is explained in Section IV). As such state is varying from the input data, the forecast uncertainty also changes with the time-series data. Then, this change could be modelled as a Markov decision process (MDP). {{In this way, we innovatively introduce the reinforcement learning algorithm to assist the LSTM in order to obtain the forecast uncertainty.}} The reason for selecting reinforcement learning is that it is a powerful technique to improve the artificial target by the autonomous learning without using any prior knowledge \cite{Guokai}. {{Here, the target is to obtain the effective probabilistic forecast, and the reinforcement learning is used to learn the forecast uncertainty by observing cell states of the LSTM.}}

{{The contributions of this paper are shown as follows:}}

1) {{A reinforcement learning assisted deep learning framework is proposed for probabilistic EVCS charging power forecasting.}} To the authors' best knowledge, this is the first paper that uses the reinforcement learning algorithm to obtain forecast uncertainty in the field of EV charging power forecasting.

2) We model the variation of LSTM cell state as a MDP, which is solved by proximal policy optimization (PPO) to obtain the forecast uncertainty. In this case, the expected value of the EVCS charging power is forecasted by the LSTM, while the forecast uncertainty are yielded by the PPO.

3) An adaptive exploration PPO (AePPO) is further proposed. This would help adaptively balance the exploration and the exploitation during the training of PPO, which could improve its performance and help prevent it from getting trapped into local optima.

4) {{A data transformer method is developed to obtain the information of EVCS from distributed charger recordings, i.e., charging sessions for the LSTM training.}} It aggregates the charging sessions and transforms them into the time-series format that includes the information of charging process, i.e., the charging power, utilization time and demand satisfaction rate.

The remainder of this paper is organized as follows. Section II presents our proposed framework for EVCS charging power forecasting. In Section III, a data transformer method is proposed to preprocess the charging data. Section IV introduces the LSTM, the MDP model of the cell state variation and the original PPO. In Section V, the AePPO is proposed, and the case studies are conducted in Section VI. Finally, Section VII concludes this paper.

\section{Framework}\label{sec:4}
\subsection{Problem Formulation}
{{The target of our studied probabilistic forecast problem is modelled by following formulations.}}
\begin{equation}
	\widetilde{F}_{t+1}=\hat{E}_{t+1} + \epsilon \label{eq:problem_start}
\end{equation}
\begin{equation}
	\epsilon \sim \mathcal{N}(0,\delta_{t+1})
\end{equation}
\begin{equation}	
	\hat{E}_{t+1}=f(X_t)
\end{equation}
\begin{equation}
	\delta_{t+1} = g(f,X_t)\label{eq:action}
\end{equation}
{{where $\widetilde{F}_{t+1}$ is the probabilistic forecast of EVCS charging power at time $t+1$. $\hat{E}_{t+1}$ is the expectation of the forecast values, which is obtained by the point forecast function $f(\cdot)$, as shown in (3) \cite{zhu_deep_2017}.}} $\epsilon$ represents the noise, which stands for the uncertainty between the forecast and real values. {{In this paper, it is assumed that the $\epsilon$ is Gaussian distributed with variance $\delta_{t+1}$, which is obtained by the variance forecast function $g(\cdot)$, presented in (4).}} It has been proved that the time series models based on the Gaussian distribution assumption can be applied to obtain a satisfactory performance \cite{6665108}. Note that $f(\cdot)$ and $g(\cdot)$ represent the LSTM and our proposed AePPO, respectively. In addition, $X_t=\{x_{t-T},...,x_{t-1},x_t\}$ is the vector of input features of the two functions. It stands for the features of $T$ timestamps before $t+1$, whereas $x_t \sim \mathbf{N}^{N_f}$ ($\mathbf{N}^{N_f}$ denotes the $N_f$ dimensional integer space).

Generally, $f(\cdot)$ provides a point forecast which represents the mean value of our target \cite{khosravi_comprehensive_2011}. However, the obtained results might be unreliable since its inherent parameters are learned by the stochastic gradient descent optimization method such as Adam \cite{zhang_deep-learning-based_2021} and its input data $X_t$ is related to the stochastic situation, i.e., the charging behavior of EV users. {{Therefore, it is necessary to quantify the uncertainty of $f(\cdot)$ and $X_t$, and the introduction of the noise $\epsilon$ is one of the most effective approaches \cite{6665108, sun_using_2020}.}}

{{Usually, the $\delta_{t+1}$ is forecasted through statistical indicators, which requires a massive repeated running of $f(X_t)$ and thus leads to a huge computation complexity \cite{zhu_deep_2017}. Consequently, it is expected to design a function $g(\cdot)$ to forecast the variance $\delta_{t+1}$ without the repeated running. The $g(\cdot)$ should have the ability to capture the uncertainty of $f(\cdot)$ and $X_t$, which means the input of $g(\cdot)$ has to contain the characteristics of both point forecast model and data. Compared with other deep learning algorithms, such as ANN and conventional neural network, the LSTM owns a unique structural parameter, i.e., the cell state, which is the core parameter of LSTM and is also determined by the $X_t$. Therefore, the uncertainties coming from both $f(\cdot)$ and $X_t$ would make changes to the value of cell state, which could be captured by the $g(\cdot)$ to forecast $\delta_{t+1}$. Besides, since the value of cell state is affected by its previous value, the change is modelled as a MDP in this paper. Therefore, the cell state is denoted as the state of the MDP, and the $\delta_{t+1}$ is defined as the action. In this way, a reinforcement learning model, AePPO, is developed as $g(\cdot)$ to observe the cell state and generate the action, i.e., the $\delta_{t+1}$. Generally, when the LSTM is applied to produce $\hat{E}_{t+1}$, the cell state would be produced, and the AePPO would capture it to determine $\delta_{t+1}$. Then, the probabilistic forecast can be constructed by $\hat{E}_{t+1}$ and $\delta_{t+1}$.}}

\subsection{The Probabilistic Forecast Framework of EVCS Charging Power}
\begin{figure}[htp]
	\centering
 	\includegraphics[width=3.2in]{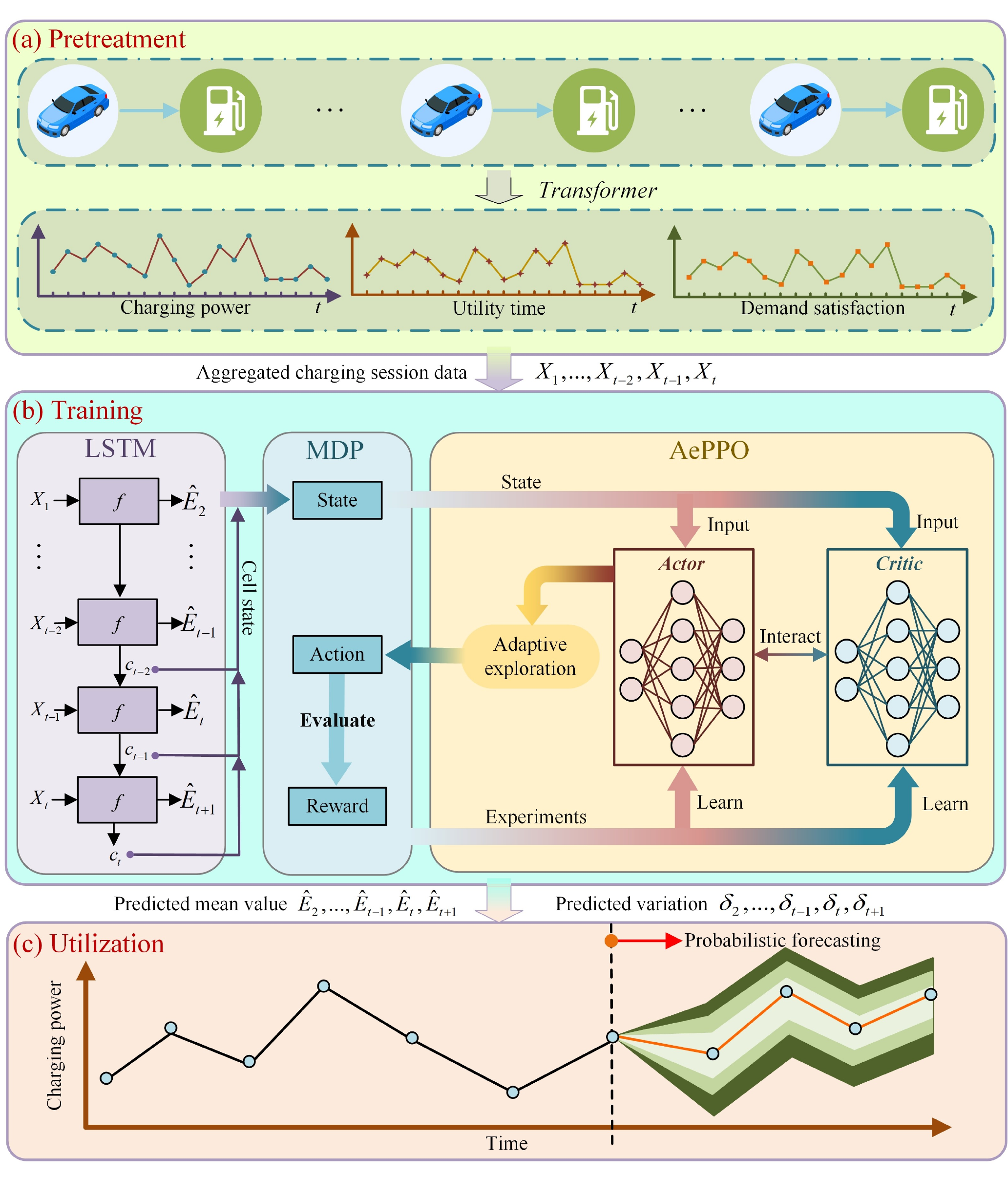}
	\caption{The overall structure of the proposed forecast framework.}\label{fig:Framework}
\end{figure}

As discussed above, the LSTM and AePPO are trained and utilized to obtain the mean and variance values for obtaining the predicted probabilistic distribution of EVCS charging power. In order to implement this procedure, we propose a probabilistic forecast framework, as shown in Fig. \ref{fig:Framework}. It contains three parts, preprocessing, training, and utilization. They are presented by the three sub-figures in Fig. \ref{fig:Framework}, which are denoted as (a), (b) and (c), respectively. 

As illustrated in Fig. \ref{fig:Framework}(a), an EVCS has numerous chargers which record the information about charging power in the period between the arrival and departure of an EV user \cite{buzna_ensemble_2021}. Since all chargers may work simultaneously, the charging information of EVCS is obtained by recordings of chargers. In this paper, such recording is termed as the charging session. It consists of the period during the process of EV charging, i.e., the arrival and departure time of the EV, and the related data about the energy information, for instance, the demand and remaining energy of the corresponding EV. Note that one charging session only contains partial charging information of EVCS and the period between different charging sessions may overlap. The information in charging session during the overlap period should be gathered to derive the data of EVCS for LSTM training. {{In this case, the charging session data should be pre-processed by aggregating the information and transforming them into the time-series format with different features, such as the charging power, utility time and demand satisfaction rate.}}

{{During the training process shown in Fig. \ref{fig:Framework}(b), the LSTM and AePPO are separately trained. At first, the EVCS data is used to train LSTM. Then, for the well-trained LSTM, its variation of cell state is modeled as a MDP. By doing so, AePPO could produce the action and interact with the LSTM according to the state. Afterwards, the reward is calculated based on the state and action, which is used for the training of AePPO.}}

Finally, the last part of this framework is the utilization, as shown in Fig. \ref{fig:Framework}(c). The LSTM and AePPO models are utilized here. That is, the LSTM provides the mean value of probabilistic forecast distribution $\hat{E}_{t+1}$, while AePPO is used to determine the variation $\delta_{t+1}$ based on the cell state of LSTM. In this way, the predicted probabilistic distribution $\widetilde{F}_{t+1}$ could be obtained by Eq. (\ref{eq:problem_start}).

\section{Data Transformer Method}\label{sec:pre}
In this section, the data transformer method is introduced to extract the charging information of EVCS from charging sessions, including the delivery energy, utility time and demand satisfaction rate. Then, they are aggregated into the time-series format for the training of LSTM.

\begin{figure}[h]
	\centering
	\includegraphics[width=2.8in]{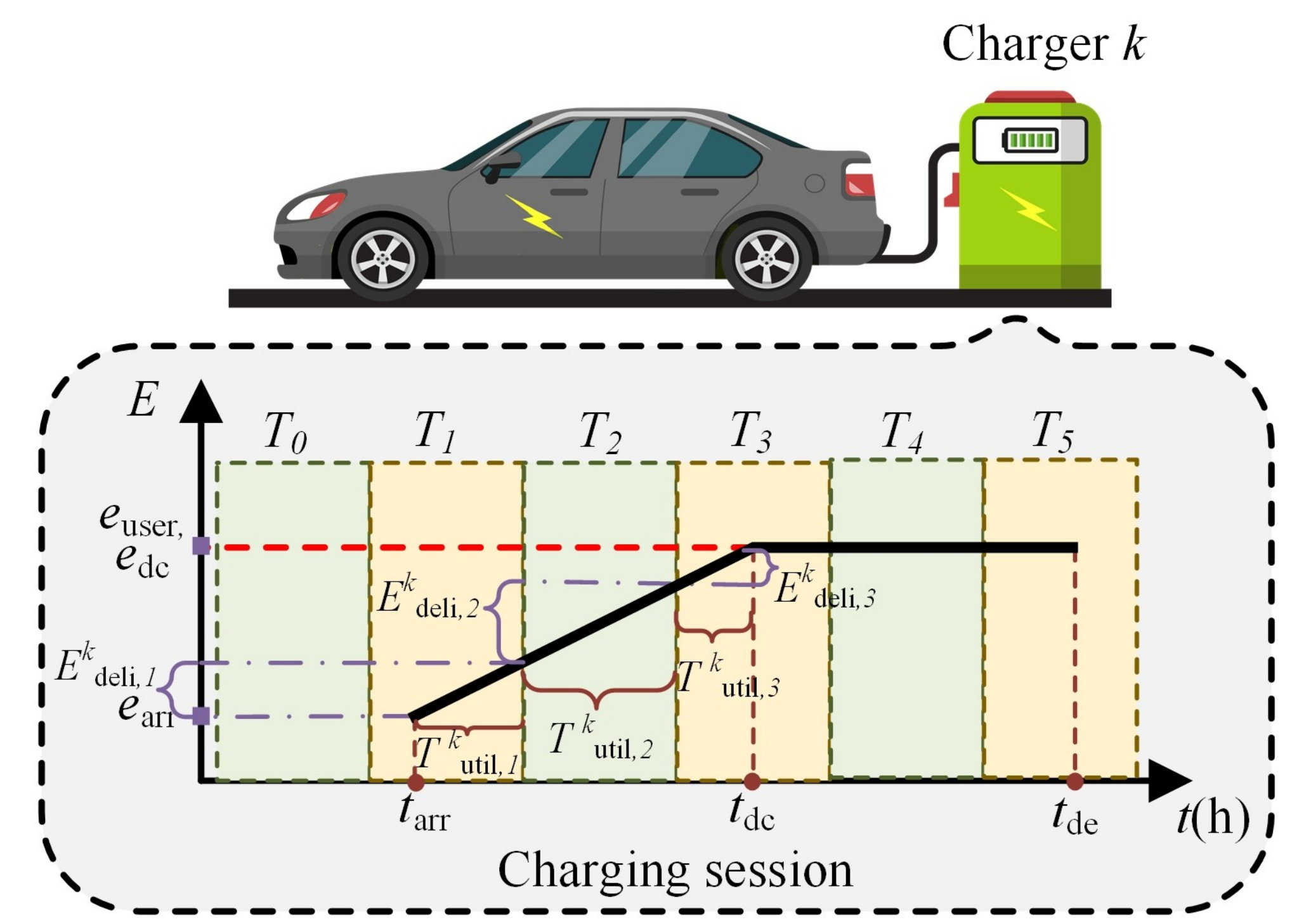}
	\caption{The charging session recorded by charger $k$. }\label{fig:Session}
\end{figure}

Usually, the charging session data is composed by a series of time and energy information which is recorded during the charging process of the EV. An example of a charging session recorded by the $k$th charger is illustrated in Fig. \ref{fig:Session}. The time information includes $t_{\text{arr}}$, $t_{\text{dc}}$ and $t_{\text{de}}$, which denote the time when EV is connected to the charger, done charging and departure. {{Besides, the energy information is also recorded, including $e_{\text{arr}}$, $e_{\text{dc}}$ and $e_{\text{de}}$, which represent the remaining energy in the EV battery at $t_{\text{arr}}$, $t_{\text{dc}}$ and $t_{\text{de}}$, respectively.}} Furthermore, the charging session also contains the user demand, termed as $e_{\text{user}}$. {{Note that expect the above information, the energy $e_t$ at the end of timestamp $t$ is also recorded.}}

However, it is difficult to use the original charging session data for the forecasting of EVCS charging power. The charging session focuses on a single EV, which only contains partial information of the EVCS since numerous EVs may charge at the same time. 

Therefore, the charging session is manually split by timestamps, and each of the timestamps represents 1 hour, as illustrated in Fig. \ref{fig:Session}. Three features are defined to summarize the charging information in each timestamp $t$, i.e., the delivery energy $ E^k_{\text{deli},t}$, the demand satisfaction rate $ D^k_{t} $ and the utilization time $ T^k_{\text{util},t} $. They are formulated as follows.
\begin{equation}
E^k_{\text{deli},t}=\left\{
	\begin{array}{lr}
		e_t-e_{\text{arr}}, & t_{\text{arr}}<t< t_{\text{arr}}+1\\
		e_t-e_{t-1}, & t_{\text{arr}}+1\leq t<t_{\text{dc}}\\
		0,& t_{\text{dc}}\leq t<t_{\text{de}}
	\end{array}
\right.\label{eq:data_start}
\end{equation}
\begin{equation}
	T^k_{\text{util},t}=\left\{
	\begin{array}{lr}
		t-t_{\text{arr}}, & t_{\text{arr}}<t< t_{\text{arr}}+1\\
		1, & t_{\text{arr}}+1\leq t<t_{\text{de}}
	\end{array}
	\right.
\end{equation}
\begin{equation}
	D^k_{t}=\frac{E^k_{\text{deli},t}}{e_{\text{user}}-e_{\text{arr}}} \times 100\% \label{eq:data_end}
\end{equation}
{{where $k$ indicates the charger. Examples of these features are shown in Fig. \ref{fig:Session}, where the charging session is split into 6 timestamps.}}

{{Then, considering $N_c$ chargers are set in the EVCS, the features split by timestamp are aggregated by the following equations:}}
\begin{equation}
E_t=\sum_{k=1}^{N_c}E^k_{\text{deli},t}, \quad T_t=\sum_{k=1}^{N_c}{T^k_{\text{util},t}},\quad D_t=\sum _{k=1}^{N_c}D^k_{t}
\end{equation}

{{After aggregation, the three features are used for the training of LSTM. They could summarize the information of time and energy contained in the charging session data. On the one hand, the delivery energy measures the service quality of the EVCS in terms of time. On the other hand, the time information of the charging process is summarized by the demand satisfaction rate and utilization time. In this way, they are selected as the features for the training of LSTM, aiming to promote its performance. Therefore, the number of features $N_f$ is 3 in this paper, and the input feature vector at timestamp $t$ is denoted by $x_t=[E_t,T_t,D_t]$.}}

\section{Reinforcement Learning Assisted Deep Learning Algorithm}\label{sec:2}
This section presents the reinforcement learning assisted deep learning algorithm. In the first subsection, the LSTM is introduced to forecast the mean value of charging power with the aggregated time-series data. Then, the variation of the LSTM cell state is modeled as a MDP in the second subsection. In the last subsection, the PPO is used to solve the MDP, and obtain the variance of the forecasts.
\subsection{Long Short Term Memory}
{{In this paper, in order to obtain the mean value of EVCS charging power for the forecast of probabilistic distribution, the LSTM is applied. {{Compared with other deep learning algorithms, such as the fully connected network, convolutional neural network and vanilla recurrent neural network, LSTM is more capable of learning the long-term dependencies inherent to the time-series data and would not suffer from the vanishing gradients \cite{sun_using_2020}}}. Besides,  Overall, the LSTM is formulated as follows.
\begin{equation}
y_{t+1}=A\left(X_t,y_t,c_t\right)
\end{equation}
where $y_{t+1}$ represents the output of LSTM at time $t+1$, which stands for the predicted mean value of EVCS charging power. The LSTM model A takes $X_t$, $y_t$ and $c_t$ as inputs, which denote the input data, the output and the cell state of LSTM at time $t$, respectively.}}

{{For clarity, the formulations of the LSTM is referred as follows.}}
\begin{eqnarray}
&i_{t+1}=\sigma\left(W_i\cdot\left[y_t,X_t\right]+b_i\right) \label{eq:lstm_start} \\ 
&f_{t+1}=\sigma\left(W_f\cdot\left[y_t,X_t\right]+b_f\right)\\
&o_{t+1}=\sigma\left(W_o\cdot\left[y_t,X_t\right]+b_o\right)	\\
&{\tilde{c}}_{t+1}=tanh\left(W_c\cdot\left[y_t+X_t\right]+b_c\right)\\
&c_{t+1}=f_{t+1}\odot c_t+i_{t+1}\odot{\tilde{c}}_{t+1}	\\
&y_{t+1}=o_{t+1}\odot tanh\left(c_{t+1}\right)	\label{eq:lstm_end}
\end{eqnarray}
where $\left[W_i,W_f,W_o,W_c\right]$ are the learnable weight metrics and $\left[b_i,b_f,b_o,b_c\right]$ are the learnable bias metric. $\sigma$, \textit{tanh} are the sigmoid and hyperbolic tangent activation functions. Their output range is $(0,1)$. $\odot$ indicates the elementwise multiplication. $i_{t+1}, f_{t+1}, o_{t+1}$ stand for the input, forget and output gate, respectively. ${\tilde{c}}_{t+1}$ represents the candidate information of the inputs in $c_{t+1}$, i.e., the cell state.

It can be seen from Eqs. (14) and (15) that the value of the cell state highly influences $y_{t+1}$. Besides, these equations also show that the cell state is determined by the learnable parameters and the input data $X_t$, and they are affected by the uncertainty from the model and user’s behaviour, respectively. Wherein, the uncertainty of the LSTM model comes from the learnable parameters, which are determined by the stochastic gradient descent optimization method. In addition, since $X_t$ is aggregated by the charging session data, it is uncertain as well because of correlations with the behaviour of users.

\subsection{The Modeling of LSTM Cell State Variation}
As described above, the input of LSTM $X_t$ is the time-series format, which reflects the behavior of the users. Besides, because of the recurrent structure of the LSTM, the cell state is influenced by its previous value. 
{{Therefore, the variation of the LSTM cell state is caused by the uncertainty that comes from both the model and the data. In this case, the variation of the cell state should be modelled to extract the variance of the probabilistic forecast distribution from the LSTM cell state.}} {{Since the cell state is determined by its previous state, its variation can be modelled as a MDP.}} Normally, the MDP is represented by a tuple $\langle\mathcal{S},\mathcal{A},\mathcal{P},\mathcal{R}\rangle$. $\mathcal{S}$ is the state space that stands for all possible states of the environment. $\mathcal{A}$ represents the action space, i.e., the agent interacts with the environment to produce action $a\in \mathcal{A}$ for the guidance of state transition. $\mathcal{P}=\{p(s_{t+1}|s_t, a_t)\}$ stands for the set of transition probability, and $\mathcal{R}=r(s_t,a_t), \mathcal{R}\in\mathbb{R}; \mathcal{S\times A}\rightarrow \mathbb{R}$ depends on the state and action and is termed as the reward. In our problem, the MDP can be formulated as follows:

\textit{1) Environment}: The environment produces the state $s_t$ and computes the reward $r_t$ based on the action $a_t$ of the agent. In this paper, the variance of the probabilistic forecast distribution is obtained from the variation of the cell state, which is the most representative parameter of the LSTM. {{The definition requires a premise, namely, the LSTM model is well-trained and its learnable parameters are fixed. The promise ensures the decisive position of the cell state in the model, therefore, a well-trained LSTM is defined as the environment of MDP.}}

\textit{2) Agent}: Here, the agent stands for a policy to obtain the action from observing the state. Besides, the agent could imply the self-learning from its experience to gradually improve the policy. 

\textit{3) State}: {{The definition of the state $s_t$ at time $t$ should follow two criteria. First, the state should be the most representative parameter of the environment. Second, the state should only be related to the previous state and not affected by the next state, namely, the non-aftereffect property. According to the calculation process of LSTM shown in Eqs. (\ref{eq:lstm_start})$\sim$(\ref{eq:lstm_end}), the most suitable parameter is the cell state, which not only determines the output of LSTM but also be obtained by the cell state of the previous time. Therefore, $s_t$ is set as $c_t$ and the state space equals to all the possible value of $c_t$.}}

\textit{4) Action}: As the uncertainty of the model and input data are contained in $s_t$, it is expected that the agent can capture this uncertainty to produce the variance of the forecast distribution, i.e., the action $a_t = \delta_{t}$.

\textit{5) Reward}: The reward $r(s_t,a_t)$ indicates the evaluation index of $a_t$ on the basis of the $s_t$. It gives feedback to the agent about the performance of its action, which could be used to update the agent. Since our target is to obtain the probabilistic distribution of future charging power, which is constructed by the output of the environment, i.e., LSTM and the agent's action. In our case, the reward is designed to evaluate the performance of the predicted distribution.

To quantify it, the continuous ranked probability score (CRPS) is introduced to measure of performance for probabilistic forecasts, i.e., the accuracy of the predicted probabilistic distribution \cite{CRPSscore}. The details of CRPS calculation are given in the Appendix A.

Note that the lower CRPS represents the more accurate probabilistic forecast. Therefore, the reward is defined as $r=-\zeta \times \text{CRPS}$, which should be maximized by the agent. $\zeta$ stands for the shrinkage coefficient, which 
is set as 0.75.

\subsection{Proximal Policy Optimization}

Based on the MDP, a recent reinforcement learning algorithm, called PPO, is introduced to train the agent for generating the optimal policy $\pi(a_k|s_k)$. The PPO contains two types of deep neural networks, which are termed as \emph{actor} and \emph{critic}. The \emph{actor} works as the agent, and it produces the action by a policy $\pi$ with parameter $\theta^\pi$. On the other hand, the \emph{critic} is used to evaluate the performance of \emph{actor}, parametrized by $\theta^{Q}$. {{The \emph{critic} approximates the value function $V^{\pi_{{k}}}(s_k)$, which evaluates the value of the state $s_k$, i.e., it represents the expectation of the reward from $k$ to the end of the MDP, which can be formulated as follows.
\begin{equation}
V^{\pi_k} (s)=E_\pi\ [U_k| S_k=s] \label{eq:1}
\end{equation}
where $U_k$ is the return function, which is defined as the discounted cumulative reward from state $s_k$.
\begin{equation}
U_k (s_k)=\sum_{t=k}^{T}\gamma \times r_t\label{eq:2}
\end{equation}
where $\gamma\in\left[0,1\right]$ stands for the discount factor, $r_t$ denotes the value of reward and $T$ is the finite time horizontal of the MDP.}}

The main target of the PPO is to train the \emph{actor} and \emph{critic}, by learning from the experience tuples $\langle s_k,a_k,r_{k},s_{k+1}\rangle$. It is obtained by the interaction between the \emph{actor} and the environment. The \emph{actor} generates a Normal distribution $\mathcal{N}(a_{k,\text{mean}},a_{k,\text{var}})$ after observing $s_k$. That is, the action $a_k$ is sampled from this distribution, i.e., $a_k \sim \mathcal{N}(a_{k,\text{mean}}, a_{k,\text{var}})$. Therefore, this introduces randomness in the produced action and thus create more diverse actions when observing the same state. In this way, it leads to the diversity of rewards since they are related to the action. This mechanism is termed as the exploration of PPO, which is used to enrich the experience tuple in order to prevent the actor from getting trapped in local optima.

Then, based on the prediction of the mean value $y_k$ obtained by LSTM, the predicted probabilistic distribution $\mathcal{N}(y_k,a_k)$ is constructed. Afterwards, $r_k$ could be calculated by the reward function to evaluate $\mathcal{N}(y_k,a_k)$, and the state at the next time $s_{k+1}$ is also restored. {{With the experience tuple, the parameters of both networks are updated by following equations.}}
\begin{eqnarray}
&\theta^\pi_{k+1} = \theta^\pi_{k}+\eta_{\pi}\nabla_{s,a \sim \pi_{k}}\mathcal{L}^{\text{CLIP}}\\ \label{eq:ppo_start}
&\theta^{Q}_{k+1} = \theta^{Q}_{k}+\eta_{Q}\nabla_{\theta^{{Q}_{k}}}\mathcal{L}^\text{Q}
\end{eqnarray}
where $\eta_{\pi}$ and $\eta_{Q}$ denote the learning rates of \emph{actor} and \emph{critic}, respectively.
$\mathcal{L}^{\text{CLIP}}$ and $\mathcal{L}^\text{Q}$ stand for the loss functions of the two networks, which are given by
\begin{equation}
\begin{split}
\mathcal{L}^{\text{CLIP}}=\mathbb{E}_{s,a \sim \pi_k}\bigg[\min (\frac{\pi_k(a \mid s)}{\pi_{k-1}(a \mid s)} A^{\pi_{{k}}}_{s, a},\\
\text{clip}(\frac{\pi_k(a \mid s)}{\pi_{k-1}(a \mid s)},1-\epsilon,1+\epsilon) A^{\pi_{{k}}}_{s, a})\bigg]\\
\end{split}\label{eq:actor_loss}
\end{equation}
\begin{equation}
\mathcal{L}^\text{Q}=\mathbb{E}_{s,a \sim \pi_k}\left[(\gamma V^{\pi_{{k}}}(s_{k+1}) + r(s_k, a_k) -V^{\pi_{{k}}}(s_k))^2\right]
\end{equation}
where $\pi_{k-1}(a \mid s)$ denotes the policy of $(k-1)$th iteration, and $\gamma$ is the discount rate. $\text{clip}(t,t_{\text{min}},t_{\text{max}})$ is the clip function. It returns $t_{\text{max}}$ if $t>t_{\text{max}}$, and $t_{\text{min}}$ if $t<t_{\text{min}}$. $A^{\pi_{{k}}}_{s, a}$ is the advantage of action $a$ under state $s$, which is represented by the difference between $a$ and the averaged performance of the \emph{actor}, formulated as follows.
\begin{equation}
A^{\pi_{{k}}}_{s, a} = Q(s_k,a_k) -V^{\pi_{{k}}}(s_k)
\end{equation}

{{The Q-function $Q(s_k,a_k)$ evaluates the value of executing action $a_k$ at state $s_k$, i.e., it represents the expectation of the reward from $s_k$ to $s_T$ after selecting the action $a_k$. The formulation of the $Q$ is shown below.
\begin{equation}
Q\left(s,a\right)=E_\pi\left[U_k | S_k=s,A_k=a\right] \label{eq:5}
\end{equation}}}

{{Then, the $Q(s_k, a_k)$ is calculated by the following equation in the PPO algorithm.
\begin{equation}
Q(s_k,a_k) = r(s_k,a_k) + \sum_{i=1}^{T}\gamma^{T-i}r(s_{k-i},a_{k-i}) \label{eq:ppo_end}
\end{equation}}}

\section{Adaptive Exploration Proximal Policy Optimization}
The \emph{actor} of PPO will generate $a_{k,\text{mean}}$ and $a_{k,\text{var}}$, based on the state $s_k$. As discussed in the above section, the action $a_k$ is sampled from the distribution $\mathcal{N}(a_{k,\text{mean}},a_{k,\text{var}})$.
The value of $a_{k, \text{var}}$ determines the exploration of PPO. With the same $a_{k,\text{mean}}$, a larger $a_{k,\text{var}}$ represents a wider distribution, which brings higher degree of exploration and more diverse experiences.
On the contrary, the smaller $a_{k, \text{var}}$ cannot provide a richer experience because the action distribution is more concentrated, and the probability of $a_k$ deviating from $a_{k, \text{mean}}$ is lower. Under this situation, the PPO trains its \emph{actor} and \emph{critic} according to its previously generated experience, which is termed as the process of exploitation.

Note that exploration and exploitation are usually contradictory. Ideally, the reinforcement learning algorithm focuses on its exploration in the early stage of the training to obtain sufficient experiences. However, if the algorithm keeps a high exploration, it is hard to be convergent because of its high randomness. In this case, the exploration will shrink according to the training. Thus, the degree of exploitation is increased. Consequently, the explored experiences should be fully utilized to help the \emph{actor} and \emph{critic} converge. However, the practical training process of PPO is different from the ideal one. Since $a_{k, \text{var}}$ of PPO is generated by the \emph{actor}, which means the degree of exploration is uncontrollable. {{$a_{k, \text{var}}$ is dependent on both the initial condition (set manually) and the inherent parameters of \emph{actor}, which is difficult to converge. Thus it may cause higher exploitation prematurely. The consequence of lacking enough exploration is the lower diversity of the experiences. In this way, the PPO may get trapped in the local optima.}}

To tackle this issue, we propose an adaptive exploration mechanism for PPO to dynamically balance its exploration and exploitation. This mechanism is termed as the adaptive exploration proximal policy optimization (AePPO). In the AePPO, the action is sampled from $\mathcal{N}(a_{k,\text{mean}}, a^{\text{AePPO}}_{k, \text{var}})$ where $a_{k,\text{mean}}$ is generated by the \emph{actor} and $a^{\text {AePPO}}_{k, \text{var}}$ is determined by our proposed adaptive exploration mechanism.

{{Using such a mechanism, the exploration of AePPO is increased in the earlier training stage and then is gradually shrunk to enhance the exploitation of the experiences.}} It is inadvisable to achieve this target by gradually reducing the value of $a^{\text {AePPO}}_{k, \text{var}}$ according to the training. This is because different actions may produce the same performance, i.e., reward, which may limit the expansion of experiences. In this case, the $a^{\text {AePPO}}_{k, \text{var}}$ should be adjusted according to the performance of the reward.
\begin{figure}[h]
	\centering
	\vspace{-0.2cm}
	\includegraphics[width=3in]{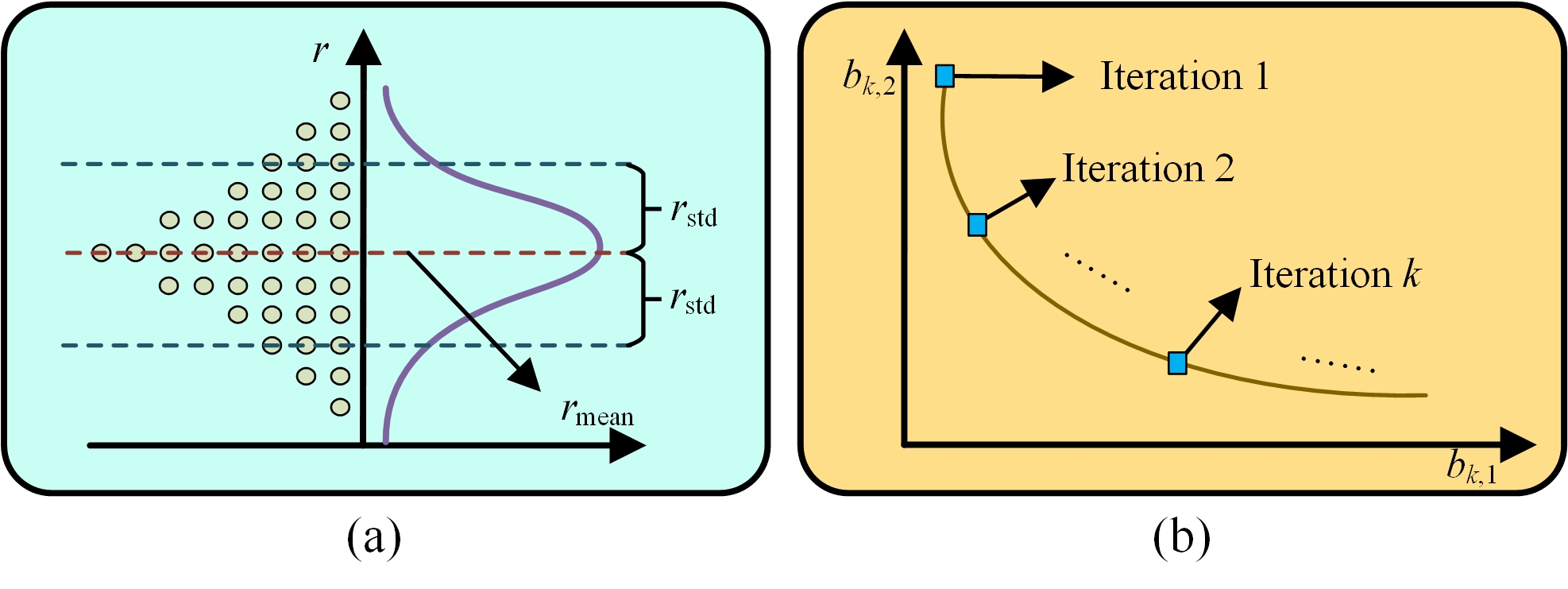}
	\caption{(a) The sampling of reward; (b) The changes of parameter with respect to iterations.}\label{fig:pso_objective}
	\vspace{-0.3cm}
\end{figure}
\begin{figure}[t]
	\centering
	\includegraphics[width=3in]{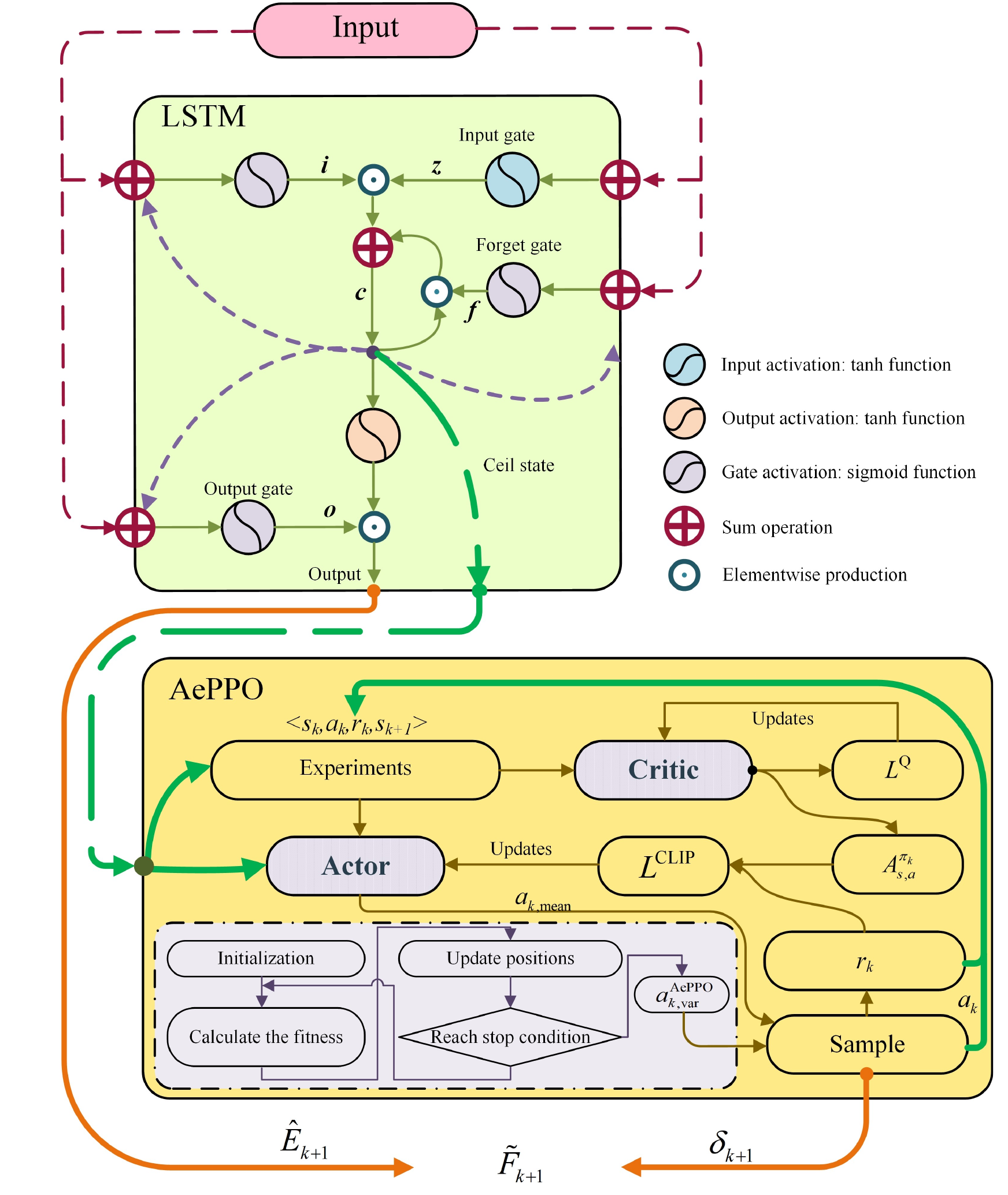}
	\caption{The interconnection between LSTM and AePPO.}\label{fig:LSTMAePPO}
	\vspace{-0.5cm}
\end{figure}

{{In PPO, the actions are sampled from $\mathcal{N}(a_{k,\text{mean}}, a^{\text {AePPO}}_{k, \text{var}})$ and the reward is determined by both the action and state, i.e., $r_k=r(s_k,a_k)$. Then, considering the Bayes' theorem, since $r_k$ is one-to-one corresponds to $(s_k, a_k)$ pair,	when the $s_k$ is fixed, $r_k$ should subject to a Normal distribution $\mathcal{N}(r^{\text{mean}}_k, r^{\text{var}}_k)$ as well, and $r^{\text{mean}}_k$ and $r^{\text{var}}_k$ represent the mean and variance of the distribution, respectively.}} As shown in Fig. \ref{fig:pso_objective}(a), the two parameters could be estimated by the following equations.
\begin{eqnarray}
&r^{\text{mean}}_k = \frac{1}{N_a}\sum_{j=1}^{N_a}r(s_k,a^i_k)\\
&r^{\text{var}}_k = \frac{1}{N_a}\sum_{j=1}^{N_a}(r(s_k,a^i_k)-r^{\text{mean}}_k)\\
&a^i_k \in (a^1_k, a^2_k,...,a^i_k,...,a^{N_a}_k)
\end{eqnarray}
where $(a^1_k, a^2_k,...,a^i_k,...,a^{N_a}_k)$ indicates the $N_a$ actions which are sampled from $\mathcal{N}(a_{k,\text{mean}},a^{\text {AePPO}}_{k, \text{var}})$. 

Note that $r^{\text{var}}_k$ represents the exploration of the PPO and it is influenced by $a^{\text {AePPO}}_{k, \text{var}}$. If $r^{\text{var}}_k$ is enlarged by changing $a^{\text {AePPO}}_{k, \text{var}}$, it indicates the diversity of experiences is enriched, i.e., the exploration ability of PPO is enhanced. Besides, since $r^{\text{mean}}_k$ evaluates the performance of PPO, the higher $r^{\text{mean}}_k$ means improved convergence as well. In this case, the focus on exploration and exploitation should be changed during the training. This process could be represented by the following equation.
\begin{equation}
f^{\text{exp}}_k = b_{k,1} r^{\text{mean}}_k + b_{k,2} r^{\text{var}}_k \label{eq:fitness}
\end{equation}
where $f^{\text{exp}}_k$ represents the degree of exploration. $b_{k,1}<1$ and $b_{k,2} = 1 - \sqrt{1-(b_{k,1}-1)^2}$ denote two parameters which represent the focus on $r^{\text{mean}}_k$ and $r^{\text{var}}_k$, respectively. As shown in Fig. \ref{fig:pso_objective}(b), the values of these parameters vary according to the training iteration $k$. In the early stage of the training, $b_{k,2}$ is larger than $b_{k,1}$, which means $f^{\text{exp}}_k$ is highly influenced by $r^{\text{var}}_k$. Then, $b_{k,1}$ is increased according to the training, and thus, $r^{\text{var}}_k$ gradually decreases. When $b_{k,1}>b_{k,2}$, the value of $f^{\text{exp}}_k$ is more influenced by $r^{\text{var}}_k$, which means the larger $f^{\text{exp}}_k$ causes higher $r^{\text{var}}_k$ and thus strengthen the exploration. On the other hand, when $b_{k,1}\leq b_{k,2}$, the raise of $f^{\text{exp}}_k$ leads to the increase of $r^{\text{mean}}_k$, the focus of $f^{\text{exp}}_k$ has changed from raising the exploration to increase the average performance of PPO, i.e., enhancing the exploitation. In this case, $f^{\text{exp}}_k$ should be maximized by adjusting the $a^{\text {AePPO}}_{k, \text{var}}$ during the training of AePPO, i.e., $-f^{\text{exp}}_k$ should be minimized.
In this paper, it is solved by the particle swarm optimization (PSO), a well-known optimization algorithm \cite{6264109}.


{{In conclusion, after the separate training, the interconnection between LSTM and proposed AePPO is described as follows.}} As illustrated in the upper side of Fig. \ref{fig:LSTMAePPO}, LSTM takes the input data, which is obtained by the data transformer and performs its internal calculation according to the Eqs. (\ref{eq:lstm_start})$\sim$(\ref{eq:lstm_end}). After that, the output of LSTM is the predicted mean value $\hat{E}_{k+1}$ of EVCS charging power. Then, the cell state of LSTM is extracted to fulfill the experience of AePPO, which serves as the state $s_k$ of AePPO. The lower side of Fig. \ref{fig:LSTMAePPO} shows the process of AePPO training iteration. The square with a yellow background in this figure represents the update mechanism of \emph{actor} and \emph{critic} according to the Eqs. (\ref{eq:ppo_start})$\sim$(\ref{eq:ppo_end}). Besides, the area with a violet background represents the procedure of adaptive exploration. 
Then, the PSO is applied to minimize $-f^{\text{exp}}_k$ and obtains $a^{\text {AePPO}}_{k, \text{var}}$, which represents the adaptive exploration of AePPO. Afterwards, the variation of predicted distribution $\delta_{k+1}$ is sampled from $\mathcal{N}(a_{k,\text{mean}}, a^{\text{AePPO}}_{k,\text{var}})$ where $a_{k,\text{mean}}$ is determined by the \emph{actor} of AePPO. Finally, based on Eqs. (\ref{eq:problem_start})$\sim$(\ref{eq:action}), the probabilistic prediction distribution of EVCS charging power is represented by $\widetilde{F}_{k+1}=\mathcal{N}(\hat{E}_{k+1},\delta_{k+1})$.
The pseudocode of LSTM-AePPO is provided in the Appendix B.

\section{Case Study}\label{sec:5}
\subsection{Data Description and Experiential Initialization}
{{In this part, we conduct a case study to verify the effectiveness of our proposed algorithm, i.e., LSTM-AePPO. The training data of the algorithm is collected from ACN dataset \cite{lee_acndata_2019}, which is an open dataset for EVCS charging researches. The charging sessions in ACN dataset are recorded from two EVCSs, one at Caltech, Pasadena, including 54 chargers and the other located in Jet Propulsion Laboratory (JPL), Lacanada, USA, containing 50 chargers. In our experiments, the charging session from 1st June 2018 to 1st June 2020 are applied to train the LSTM-AePPO, in which 25916 and 22128 charging sessions are included in the case of Caltech and JPL, respectively.}}

{{In order to demonstrate the effectiveness of LSTM-AePPO, numerous algorithms, i.e., support vector quantile regression (SVQR) \cite{he_short-term_2017}, linear quantile regression (QR) \cite{6038881}, and gradient boosting quantile regression (GBQR) \cite{8356127} are introduced for comparisons. Considering the seasonal variations of charging behavior, the performance and metric comparisons are conducted on the four seasons, respectively. Moreover, for further verifying the outperformance of our proposed AePPO, the traditional PPO is introduced to generate the $\delta_{k+1}$ from the cell state for comparisons, which is termed as LSTM-PPO.}}

{{Then, CRPS, Winkler \cite{sun_using_2020}, and Pinball \cite{wang_probabilistic_2019} are introduced as evaluation metrics. The three metrics focus on the accuracy, variation, and reliability of the forecast distribution, respectively. Their details are presented in Appendix C.}}

{{In this paper, the features of former $N_h$ hours $[x_{i-1},...,x_{i-N_h}]$ are used in the probabilistic forecasting of EVCS charging power at hour $i$. The value of $N_h$ is set as 12, and other hyperparameters of LSTM-AePPO are given in Table I. Note that the experiments are conducted on a computer with 8GB RAM, Intel i5-8265U CPU, and implemented in Python.}}
\begin{table}[h]
	\begin{center}
		{
			\caption{The hyperparameters of the proposed LSTM-AePPO.}
			\scalebox{0.83}{
				\begin{tabular}{ccc}
					\hline\hline
					Symbol& Description& Value\\
					\hline
					$\eta_{\pi}$,$\eta_{Q}$& The learning rate of \emph{actor} and \emph{critic} & 0.0001 \\
					$\eta_{lstm}$ & The learning rate of LSTM & 0.001\\
					$\epsilon$ & The clip value of \emph{actor} loss&0.1\\
					$\gamma$& The discount factor& 0.99\\
					$N_{e}$&The maximum training number of AePPO& 10000\\
					$N_{pso}$& The population of PSO & 20 \\
					$N_{var}$& The maximum iteration of PSO & 100 \\
					$e_1,e_2$& The learning factor of PSO&2 \\
					$\omega$& The inertia weight of PSO&0.7 \\
					\hline\hline
			\end{tabular}}\label{table:structure}}
	\end{center}
\end{table}
\begin{figure*}[t]
	\centering
	\includegraphics[width=5.8in]{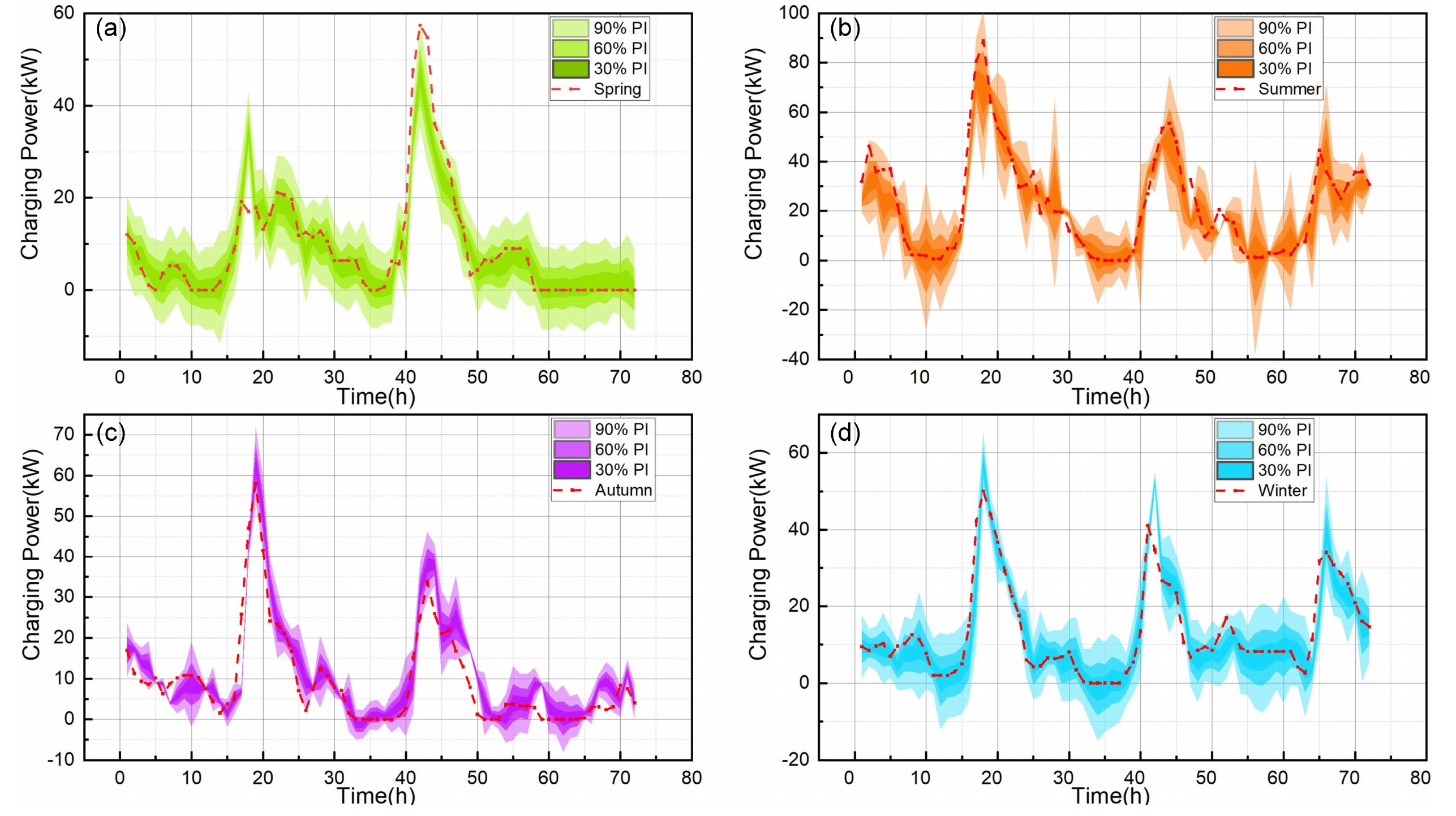}
	\caption{The probabilistic forecasting of Caltech, obtained by the proposed LSTM-AePPO framework.}\label{fig:RCProbability}
\end{figure*}
\begin{figure*}[t]
	\centering
	\includegraphics[width=5.8in]{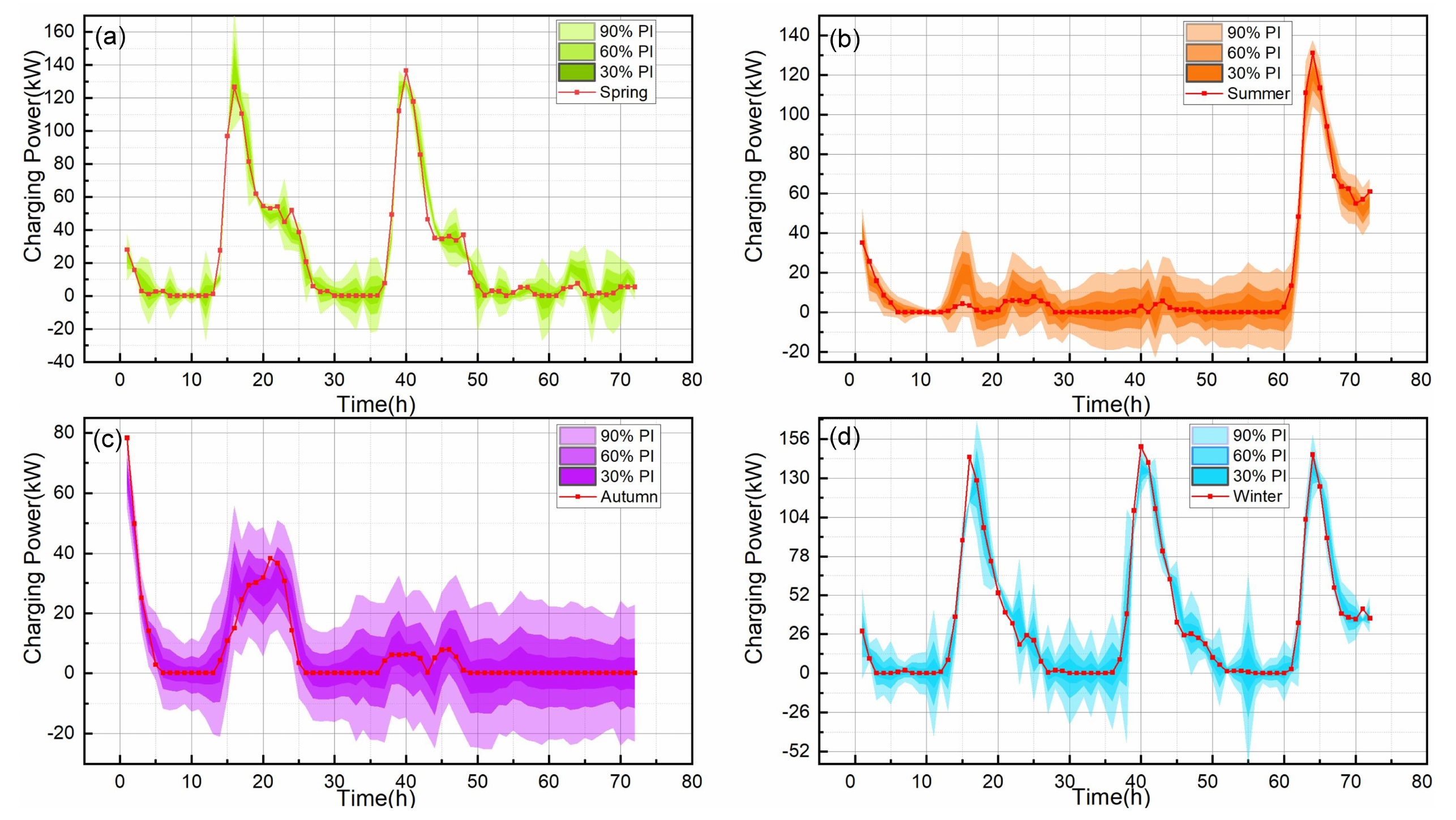}
	\caption{The probabilistic forecasting of JPL obtained by the proposed LSTM-AePPO framework.}\label{fig:RJProbability}
	\vspace{-0.3cm}
\end{figure*}
\begin{table*}[htbp]
	\centering
	\caption{The seasonal metric comparison of two sites under different PI.}
	\scalebox{0.75}[0.75]{
		\begin{tabular}{ccccccccccccc}
			\toprule
			&       &       & \multicolumn{5}{c}{Winkler}           & \multicolumn{5}{c}{Pinball} \\
			\cmidrule{4-13}          &       &       & QR    & QRSVM & GBQR  & LSTM-PPO & LSTM-AePPO & QR    & QRSVM & GBQR  & LSTM-PPO & LSTM-AePPO \\
			\midrule
			\multirow{12}[8]{*}{Caltech} & \multirow{3}[2]{*}{Spring} & 30    & 144.565  & 113.104  & 116.499  & 54.882  & \textbf{30.099} & 18.922  & 17.335  & 17.715  & 15.412 & \textbf{15.158}  \\
			&       & 60    & 85.686  & 75.070  & 73.747  & 45.429  & \textbf{21.036} & 18.023  & 16.762  & 17.095  & 18.512  & \textbf{13.311} \\
			&       & 90    & 40.323  & 39.088  & 38.607  & 34.876  & \textbf{23.966} & 16.656  & 15.783  & 15.940  & 23.966  & \textbf{14.057} \\
			\cmidrule{2-13}          & \multirow{3}[2]{*}{Summer} & 30    & 203.099  & 160.325  & 191.611  & 65.351  & \textbf{26.914} & 26.366  & \textbf{21.962}  & 24.434  & 24.424  & 23.475 \\
			&       & 60    & 99.963  & 93.854  & 102.781  & 40.923  & \textbf{16.665} & 24.387  & 20.960  & 22.590  & 23.475  & \textbf{15.835} \\
			&       & 90    & 44.193  & 45.280  & 45.503  & 27.592  & \textbf{13.074} & 21.359  & 18.935  & 20.003  & 21.815  & \textbf{18.664} \\
			\cmidrule{2-13}          & \multirow{3}[2]{*}{Autumn} & 30    & 131.399  & 83.484  & 93.647  & 37.247  & \textbf{67.858} & 18.495  & 16.616  & 18.910  & 22.154  & \textbf{14.480} \\
			&       & 60    & 76.467  & 42.663  & 47.674  & 22.865  & \textbf{35.791} & 17.788  & 16.595  & 18.434  & 15.039  & \textbf{14.683} \\
			&       & 90    & 34.970  & 28.657  & 28.326  & \textbf{18.342} & 21.510  & 16.446  & 15.818  & 16.988  & 19.850  & \textbf{14.870} \\
			\cmidrule{2-13}          & \multirow{3}[2]{*}{Winter} & 30    & 131.210  & 137.102  & 124.088  & 55.526  & \textbf{19.342} & 19.777  & 18.939  & 18.630  & 13.498  & \textbf{13.328} \\
			&       & 60    & 76.665  & 87.910  & 77.862  & 45.158  & \textbf{15.463} & 18.984  & 18.101  & 17.884  & 17.004  & \textbf{14.124} \\
			&       & 90    & 28.660  & 30.142  & 30.627  & 21.798  & \textbf{14.666} & 17.271  & 16.590  & 16.401  & 24.813  & \textbf{14.378} \\
			\midrule
			\multirow{12}[8]{*}{JPL} & \multirow{3}[2]{*}{Spring} & 30    & 367.828  & 292.015  & 278.605  & 71.544  & \textbf{27.800} & 36.793  & 30.789  & 32.484  & 23.128  & \textbf{18.072} \\
			&       & 60    & 171.225  & 157.093  & 135.559  & 26.264  & \textbf{19.512} & 34.360  & 29.845  & 30.738  & 31.896  & \textbf{20.312} \\
			&       & 90    & 54.291  & 49.069  & 49.749  & 21.784  & \textbf{18.296} & 30.001  & 26.917  & 27.372  & 46.840  & \textbf{20.760} \\
			\cmidrule{2-13}          & \multirow{3}[2]{*}{Summer} & 30    & 210.133  & 237.387  & 208.853  & 37.048  & \textbf{24.952} & 32.200  & 26.701  & 27.833  & 25.464  & \textbf{25.464} \\
			&       & 60    & 103.983  & 112.102  & 100.350  & 27.096  & \textbf{21.368} & 31.269  & 27.156  & 27.538  & 31.480  & \textbf{22.680} \\
			&       & 90    & 52.745  & 49.958  & 50.535  & 24.824  & \textbf{19.608} & 27.807  & 25.061  & 25.105  & 37.528  & \textbf{22.584} \\
			\cmidrule{2-13}          & \multirow{3}[2]{*}{Autumn} & 30    & 157.871  & 123.457  & 94.360  & 29.240  & \textbf{23.064} & 21.857  & 21.287  & 21.325  & 22.712  & \textbf{21.016} \\
			&       & 60    & 54.617  & 57.052  & 52.259  & 21.976  & \textbf{19.416} & 23.676  & 23.167  & 23.110  & 25.400  & \textbf{21.464} \\
			&       & 90    & 47.000  & 47.033  & 47.040  & 18.232  & \textbf{18.584} & 22.834  & 22.420  & 22.123  & 28.856  & \textbf{21.656} \\
			\cmidrule{2-13}          & \multirow{3}[2]{*}{Winter} & 30    & 524.566  & 569.323  & 505.956  & 38.328  & \textbf{32.440} & 45.037  & 43.167  & 40.867  & 22.680  & \textbf{22.136} \\
			&       & 60    & 258.636  & 314.319  & 265.607  & 23.416 & \textbf{23.256}  & 41.205  & 39.586  & 37.381  & 24.920  & \textbf{24.888} \\
			&       & 90    & 51.263  & 76.256  & 73.966  & 19.032  & \textbf{18.776} & 34.898  & 33.815  & 32.077  & 31.512  & \textbf{29.400} \\
			\bottomrule
		\end{tabular}%
	}
	\label{tab:addlabel}%
\end{table*}%

\subsection{The Performance of Probabilistic Forecasting Obtained by LSTM-AePPO}


{{The probabilistic forecast of EVCS charging power obtained by LSTM-AePPO are shown in Fig. \ref{fig:RCProbability} and Fig. \ref{fig:RJProbability}, which illustrate the results on Caltech and JPL cases among 3 test days (72 hours). The subfigures (a), (b), (c) and (d) represent the forecast of Spring, Summer, Autumn, and Winter, respectively. Specifically, $p\%$ prediction interval (PI) is used to denote the interval between the lower and upper quantiles at probability $p$, and the darker color denotes the PI with a smaller $p$.}} Consequently, the 30\% PI is shown with the darkest color while the 90\% PI corresponds to the lightest. Note that the real charging power is represented by the red line in these figures.

{{One of the biggest challenges of the charging power forecasting is to predict the peak, which varies according to the seasons.}} For instance, in the Spring and Autumn case of Caltech, it could be learned that the peaks emerge in around the 18th and 42nd hours, as shown in Figs. \ref{fig:RCProbability}(a) and (c). {{However, as shown in Figs. \ref{fig:RCProbability}(b) and (d), the peaks during the Summer and Winter appear 3 times, which is more frequent. Besides, as shown in Fig. \ref{fig:RJProbability}, the peaks are more irregular in the JPL case, which appear 2 times in the Spring and Autumn cases, 1 time in the Summer case and 3 times in the Winter case.
Although the peak appears irregularly, LSTM-AePPO achieves a good result.}} The fluctuation of the probabilistic forecast distribution can capture the EVCS charging power. As shown in Figs. \ref{fig:RCProbability} and \ref{fig:RJProbability}, most of the values in the red lines are covered by the 90\% PI, which means that the probabilistic forecast distribution could represent the variation of real charging power.
\subsection{Metrics Comparison Among Different Algorithms}

To verify the effectiveness of LSTM-AePPO, the Winkler, Pinball and CRPS are introduced as metrics for comparison. {{Since the values of Winkler and Pinball are related to PI, the effectiveness of our proposed algorithm is demonstrated by comparing with other algorithms under $30\%$, $60\%$ and $90\%$ PIs, as shown in Table \ref{tab:addlabel}.}}
\begin{figure}[h]
	\centering
	\includegraphics[width=3in]{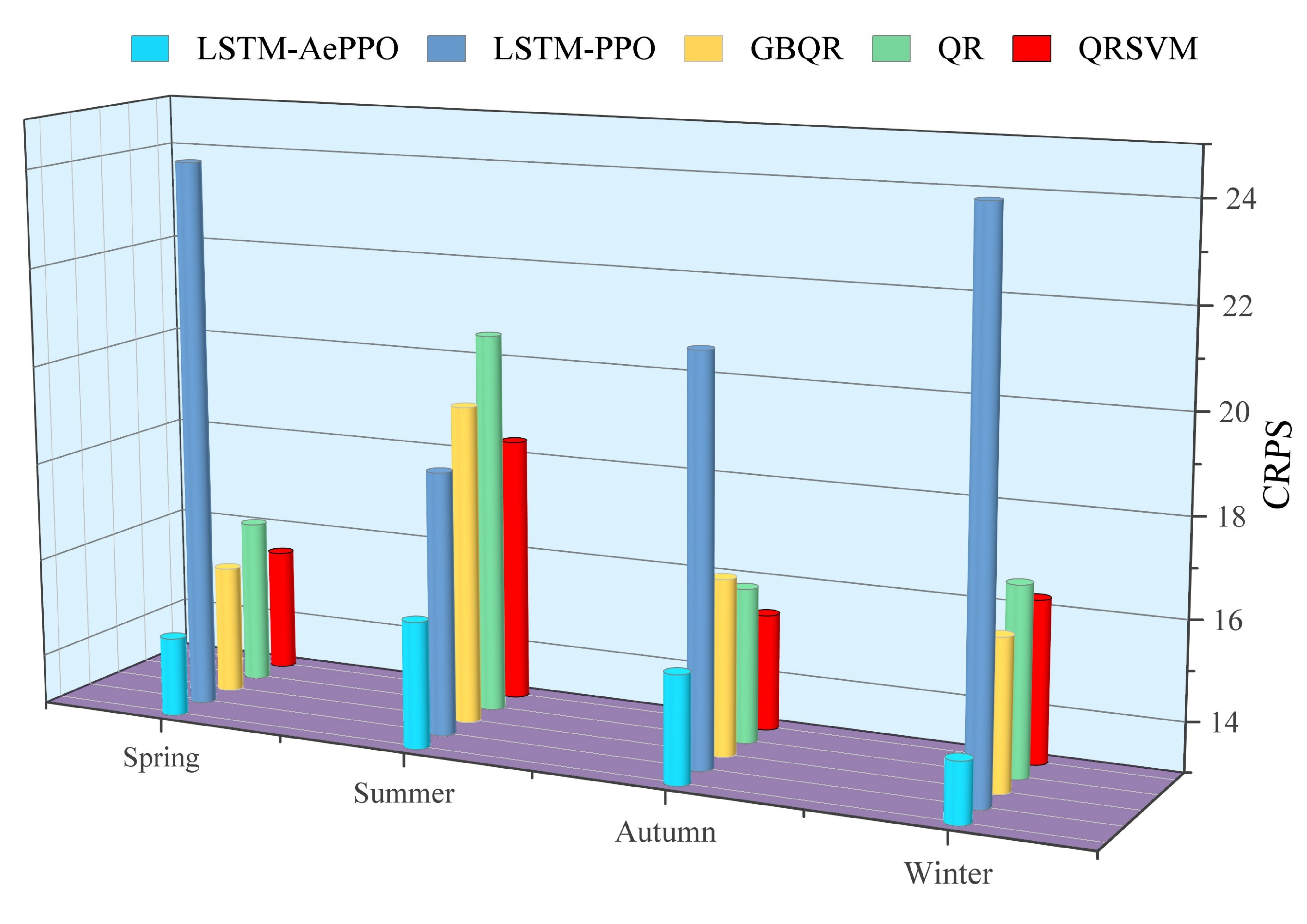}
	\caption{The comparisons of CRPS in the case of Clatech among the five mentioned algorithms.}\label{fig:CMetrics}
\end{figure}

\begin{figure}[h] 
	\centering
	\vspace{0.5cm}
	\includegraphics[width=3in]{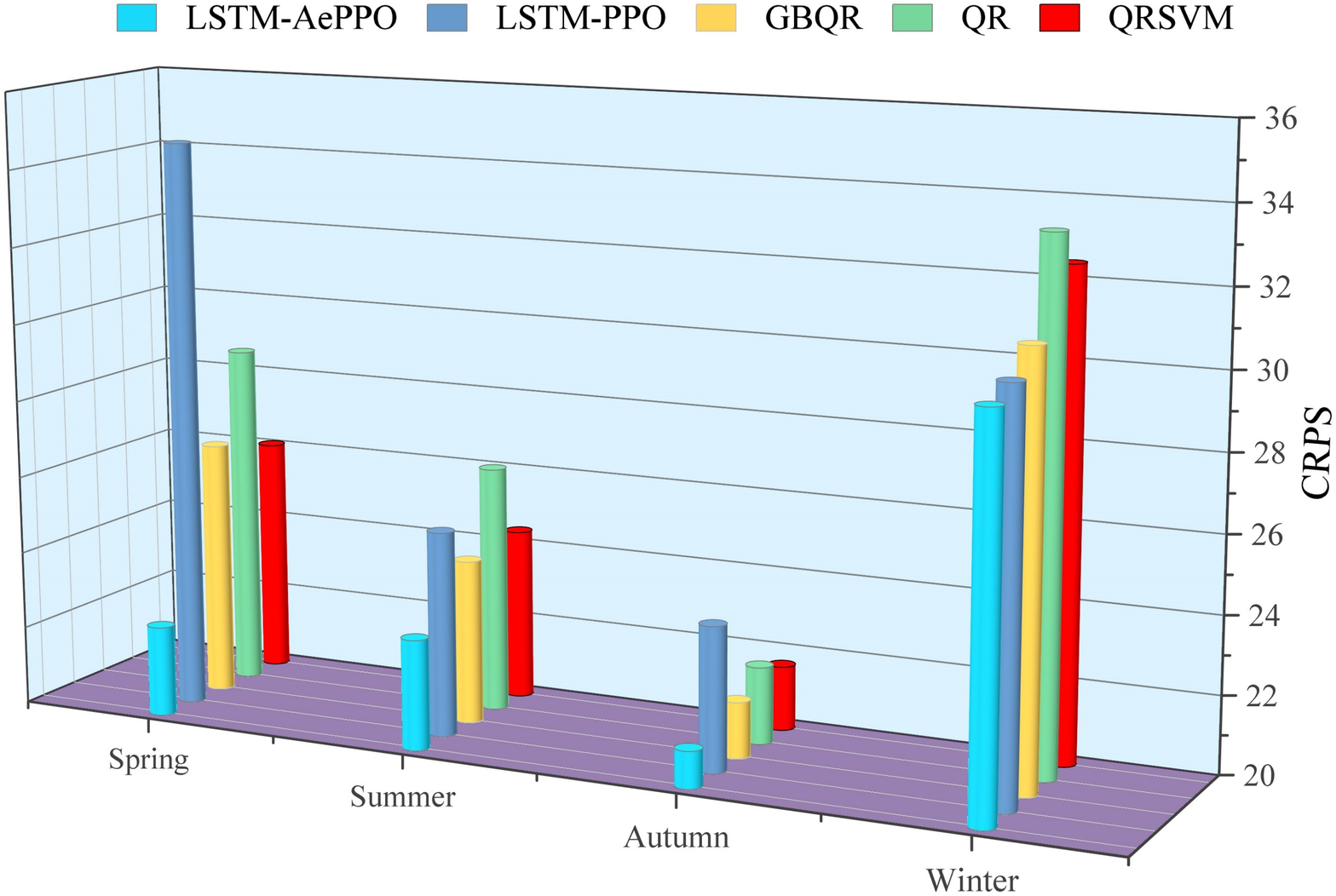}
	\caption{The comparisons of CRPS in the case of JPL among the five mentioned algorithms.}\label{fig:JMetrics}
\end{figure}

{{In this table, the best metric value of each case is marked in bold. It could be seen that the LSTM-AePPO outperforms other algorithms on Winkler and Pinball in most cases.}}
{{For example, in the Spring case of Caltech, the Winkler and Pinball values of LSTM-AePPO at 30\%, 60\% and 90\% PIs are (30.099, 15.158), (21.036, 13.311) and (23.966, 14.057), which are lower than the Winkler values of other algorithms.}} Besides, in the Summer and Autumn cases of Caltech, the LSTM-AePPO always performs the best at 60\% PI. {{Moreover, when focusing on the Winter case of Caltech, the QR, QRSVM, GBQR and LSTM-PPO also perform worse than LSTM-AePPO since their Winkle and Pinball values are larger.}} Generally, the forecast distribution obtained by LSTM-AePPO have a better variation and reliability at Caltech case.
In addition, the effectiveness is also verified comprehensively in the JPL case. As shown in the Table \ref{tab:addlabel}, the Winkler and Pinball numbers obtained by LSTM-AePPO in the JPL case are bold, which means QR, QRSVM, GBQR and LSTM-PPO are all surpassed from both the variation and reliability of the forecast distribution.



{{Note that the LSTM-AePPO does not perform the best in the following two cases, i.e., the Winkler of Caltech Autumn case at 90\% PI (21.510 for LSTM-AePPO but 18.342 for LSTM-PPO) and the Pinball of Caltech Summer case at 30\% PI (23.475 for LSTM-AePPO but 21.962 for QRSVM).}} The cause of this phenomenon may be the correlation between Winkler and Pinball to the PI. {{Therefore, as a metric that is independent of PI, the CRPS is introduced to evaluate the accuracy of the whole forecast probabilistic distribution, and the lower CRPS value indicates a more accurate forecast probabilistic distribution. The comparisons of CRPS under Caltech and JPL cases are illustrated in Fig. \ref{fig:CMetrics} and Fig. \ref{fig:JMetrics}. It can be seen from the two figures that our proposed LSTM-AePPO achieves the lowest CRPS values in all the cases, which demonstrates the accuracy of LSTM-AePPO could surpass other comparison algorithms.}} Based on the analysis of the Winkler, Pinball and CRPS values under different cases, it could be concluded that the LSTM-AePPO is more effective than QR, GBQR, QRSVM and LSTM-PPO.

\subsection{The Effectiveness of AePPO}
To illustrate the effectiveness of the proposed adaptive exploration mechanism in LSTM-AePPO, the reward curves of PPO and AePPO during the 10000 training iterations are compared in the Caltech and JPL cases, as shown in Fig. \ref{fig:CReward} and Fig. \ref{fig:JReward}. In Fig. \ref{fig:CReward}, the reward of PPO increases during the first 3000 iterations and stays steady. {{Similarly, as shown in Fig. \ref{fig:JReward}, the reward of PPO continuously raise from -20 to -7.5 in the first 1000 iterations then maintains at a relatively stable level}}. 

{{\textit{Because of applying the adaptive exploration mechanism, the AePPO would focus on exploration in the early stage, which enriches the experience while causing the backwardness of the reward value. For instance, in the case of Caltech, the reward of PPO increases from around -30 to -23 during the early 3000 iterations, while the reward value of AePPO raises from -29 to -27. The same situation also appears in the case of JPL. The reward of PPO goes up from -25 to -7 in the early training, while the corresponding reward of AePPO hardly increases. However, the exploration of PPO would decrease due to the premature convergence of its \emph{actor}, thus the diversity of its experience is not plentiful enough, which may cause the PPO falling into the local optima. As a result, the rewards of AePPO increase and surpass PPO after around the 8000th and the 5000th iteration in the case of Caltech and JPL, respectively.}}

\begin{figure}[h]
	\centering
	\includegraphics[width=2.8in]{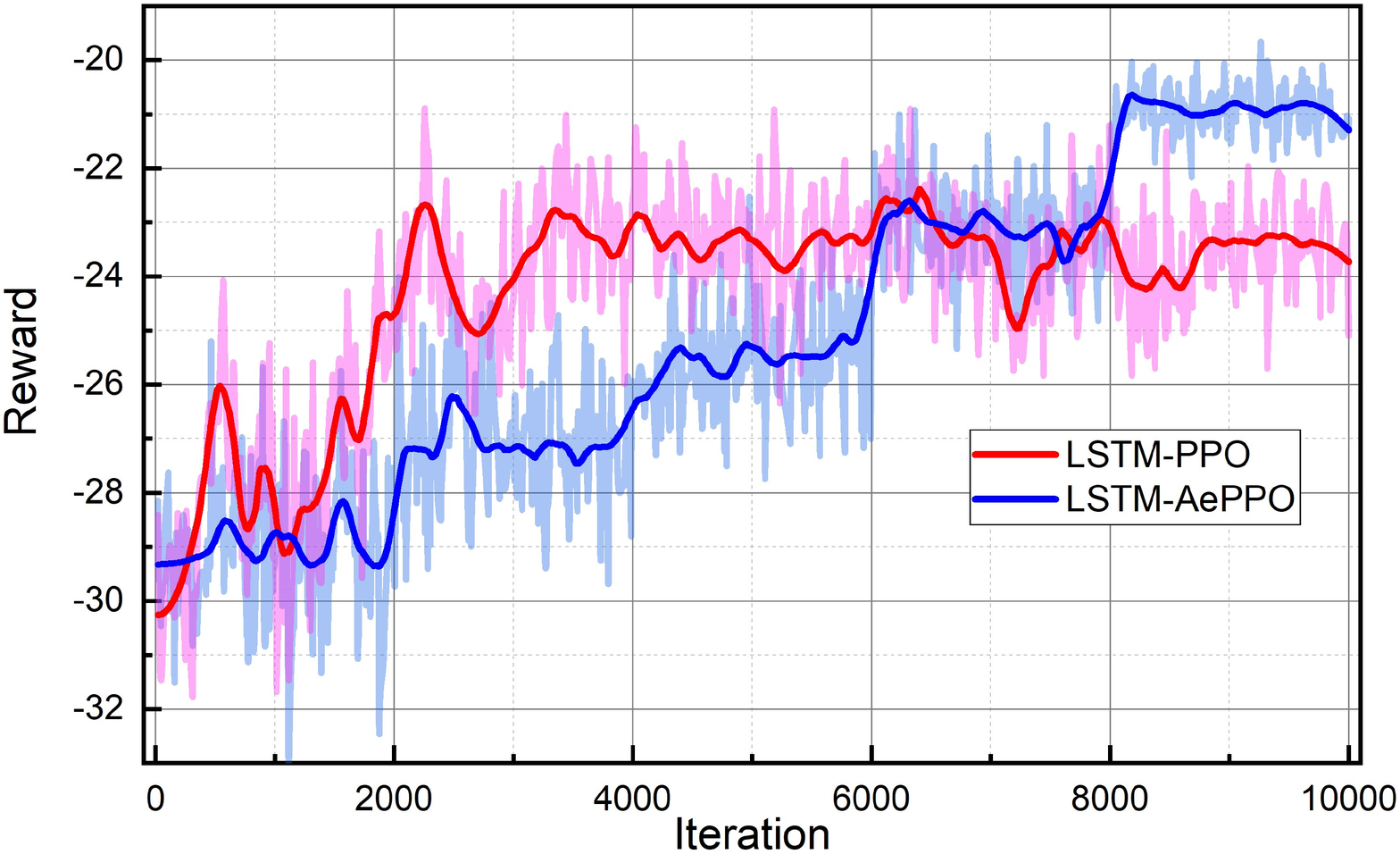}
	\caption{{{The comparison between the reward of PPO and AePPO during the training in the case of Caltech.}}}\label{fig:CReward}
\end{figure}
\begin{figure}[h]
	\centering
	\includegraphics[width=2.8in]{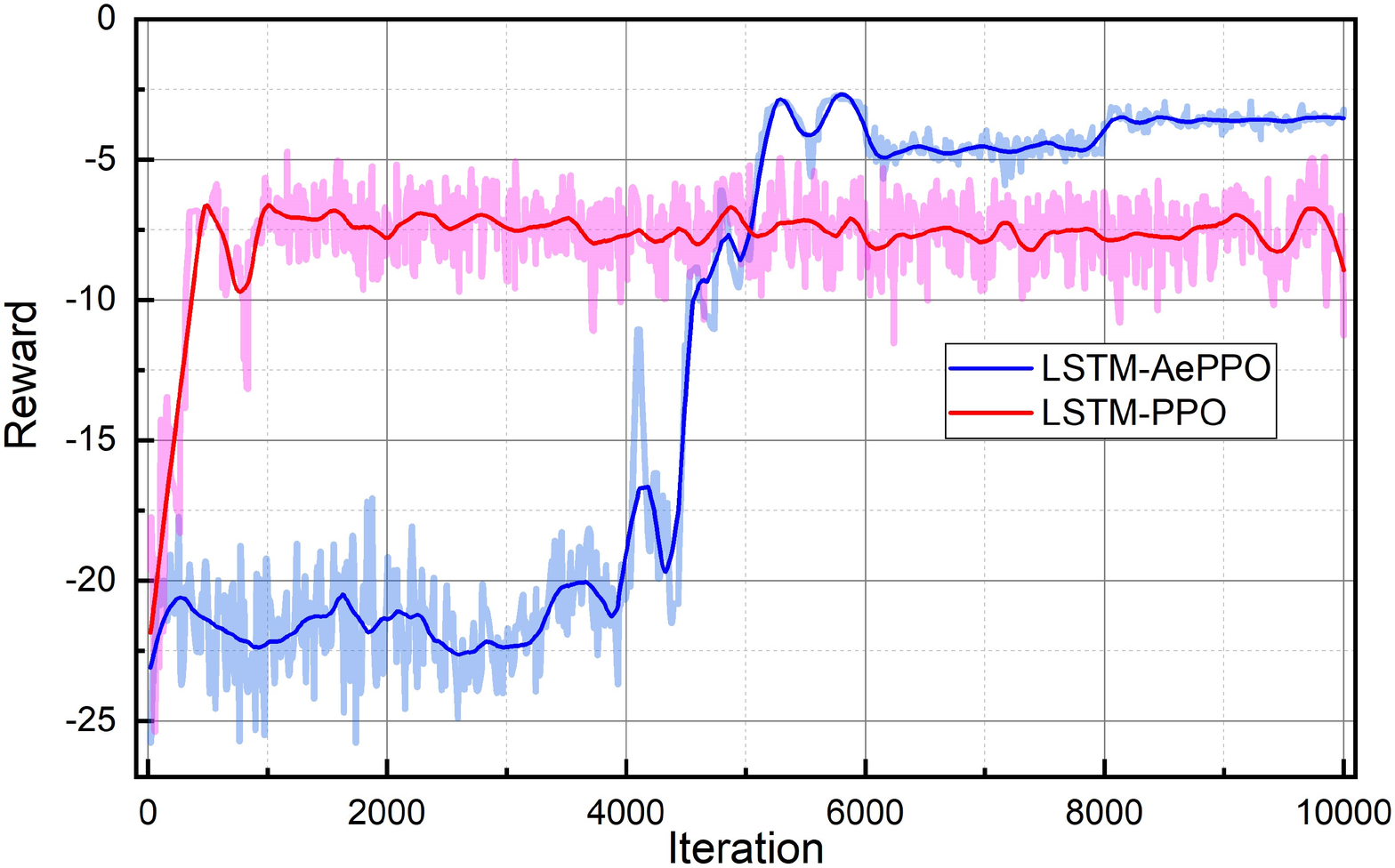}
	\caption{{{The comparison between the reward of PPO and AePPO during the training in the case of JPL.}}}\label{fig:JReward}
	\vspace{-0.2cm}
\end{figure}

Therefore, it can be seen that the adaptive exploration mechanism strengthen the accumulation of AePPO experiences in the early stage. Then, the AePPO would make full use of these experiences during the training and finally achieves a higher reward than PPO. In addition, the fluctuation of reward curves obtained by LSTM-AePPO are lower in the two cases. This illustrates that the proposed AePPO provides a more stable and superior performance than the original PPO.

\section{Conclustion}
{{This paper has proposed a reinforcement learning assisted deep learning probabilistic forecast framework for the charging power of EVCS, which contains a data transformer method and a probabilistic forecast algorithm, termed as LSTM-AePPO. After being preprocessed by the data transformer, the charging session data is used to train the LSTM, which aims to obtain the mean value of the forecast distribution. Then, the variation of the LSTM cell state is modeled as a MDP and a reinforcement learning algorithm, AePPO, is applied to solve it. In this way, the variance of the forecast distribution can be provided by the AePPO. In addition, to balance the exploration and the exploitation, an adaptive exploration mechanism is further proposed to enrich the diversity of the experiences thus preventing the premature convergence of AePPO. Finally, the case studies are conducted based on the charging session data of Caltech and JPL. The experiments show the superior performance of the proposed LSTM-AePPO by comparing with QR, QRSVM, GBQR and LSTM-PPO on the CRPS, Winkler and Pinball metrics.
Moreover, the comparison between the reward curves of PPO and AePPO indicates that the adaptive exploration mechanism is more effective in balancing the exploration and the exploitation of reinforcement learning.}}

{{Two challenges remain in our work, which are worth investigating in future work. First, the better point forecasting method should be investigated to achieve a higher prediction accuracy. {{Meanwhile, the method is also required to remain the parameters that contain the uncertainties of model and data simultaneously, such as the cell state in LSTM.}} The other is accelerating the training process of reinforcement learning algorithm, i.e., AePPO, while keeping the diversity of exploration and the convergence performance.}}

\begin{appendices}

\bibliographystyle{IEEEtran}
\bibliography{Reference}

\section{The Calculation of Continuous Ranked Probability Score}
\setcounter{equation}{0}
\renewcommand\theequation{A.\arabic{equation}}
Normally, the CRPS is defined as follows.
\begin{equation}
	\text{CRPS}(\widetilde{F},y)=\int_{-\infty}^{y} \widetilde{F}(x)^2\text{d}x+\int_{y}^\infty (1-\widetilde{F}(x))^2\text{d}x \label{CRPS_base}
\end{equation}
where $\widetilde{F}(\cdot)$ is the predicted distribution obtained by the probabilistic forecasting model and $y$ is the real value. Since the noise of our studied probabilistic forecast problem is assumed to follow the Gaussian distribution, as shown in Eqs. (\ref{eq:problem_start})$\sim$(\ref{eq:action}), Eq. (\ref{CRPS_base}) can be derived as follows.
\begin{equation}
	\begin{split}
		\text{CRPS}(\widetilde{F},y) = \delta(\frac{y-\mu}{\delta}\left(2\widetilde{F}\left(\frac{y-\mu}{\delta}\right)-1\right)+\\2\widetilde{f}\left(\frac{y-\mu}{\delta}\right)-\frac{1}{\sqrt{\pi}})
	\end{split}\label{eq:crps}
\end{equation}
where $\mu$ and $\delta$ are the mean and variance of the predicted distribution. {{The functions $\widetilde{F}(\cdot)$ and $\widetilde{f}(\cdot)$ are formulated as follows.}}
\begin{eqnarray}
	\widetilde{f}(x)=\frac{1}{\sqrt{2\pi}}e^{\left(-\frac{x^2}{2}\right)}\\
	\widetilde{F}(x)=\int_{-\infty}^{x}\widetilde{f}(t)\text{d}t
\end{eqnarray}

\section{The Pseudocode of LSTM-AePPO Training}
\begin{breakablealgorithm}
	\caption{{{The training of LSTM.}}}
	\begin{algorithmic}[1]
		\State\textbf{Input} The set of input features $\{X_n\}_{n=1}^{N}$ and the target charging power set $\{Y_n\}_{n=1}^{N}$.
		\State\textbf{Output} A well trained LSTM model $A(X_n, Y_n, c_n)$
		\State\textbf{Initialize} $\eta_{lstm}$ (The learning rate of LSTM).
		\State\textbf{Initialize} $N_{train}, N_{valid}, N_e$ (Number of samples in the training set, validation set and the number of training epochs)
		\Statex
		\For{epoch $e=1$ to $N_e$}
		\State \textbf{Initialize} $c_1$
		\State Receive $X_n$ and $Y_n$
		\State Execute the forward propagation of LSTM:
		\State $f_{n+1}\leftarrow \sigma(W_f^n\cdot [Y_{n},X_n]+b_f^n)$
		\State $i_{n+1}\leftarrow \sigma(W_i^n\cdot [Y_{n},X_n]+b_i^n)$
		\State $\tilde{c}_{n+1}\leftarrow tanh(W_c^n\cdot[Y_{n},X_n]+b_{c}^n)$
		\State $c_{n+1}\leftarrow f_{n+1}\otimes c_{n}+i_{n+1} \otimes \tilde{c}_{n+1}$
		\State $o_{n+1}\leftarrow \sigma(W_o^n[y_{n},X_t]+b_o^n)$
		\State $\hat{y}_{n+1} \leftarrow o_{n+1}\times tanh(c_{n+1})$
		\State Calculate the loss of LSTM:
		\State Receive $Y_{n+1}$
		\State $\mathcal{L}$ $\leftarrow$ $\text{MSE}(\hat{y}_{n+1},Y_{n+1})$
		\State Calculate the derivatives of the learnable parameters of LSTM ${\mathrm{d}\mathcal{L}}/{\mathrm{d}\theta^{n}}$, where $\theta^{n}$ represents $W_f^n, W_i^n, W_c^n, W_o^n, b_f^n, b_i^n, b_{c}^n$ and $b_o^n$. The details are given in the Appendix D.
		\State Update the parameters of the LSTM via Adam.
		\EndFor
		\State \Return $A(X_n, Y_n, c_n)$.
	\end{algorithmic}	
\end{breakablealgorithm}

\begin{breakablealgorithm}
	\caption{{{The training of AePPO.}}}
	\begin{algorithmic}[1]
		\State\textbf{Input} The input features set $\{X_n\}_{n=1}^{N}$ and target charging power set $\{Y_n\}_{n=1}^{N}$. The cell state produced by LSTM $\{c_n\}_{n=1}^{N}$.
		\State\textbf{Output} A well trained actor $\pi$.
		\State\textbf{Initialize} $\eta_{\pi}$, $\eta_{Q}$, $\epsilon$, $\gamma$, $N_{train}$, $N_{valid}$, $N_e$, $N_{pso}$, $N_{var}$, $e_1$, $e_2$. 
		\State \textbf{Initialize} The actor and critic network $\pi$, $V$.
		\State \textbf{Initialize} $v^j_i$, $x^j_i$ and $f^j_i$ of each particle $i,i\in[1,N_{pso}]$
		\Function{fitness}{$x_i$, $b_{j,1}$}
		\State $b_{j,2} = 1 - \sqrt{1-(b_{j,1}-1)^2}$
		\State Sample $N_a$ action from $\mathcal{N}(a_{k,\text{mean}},x_i)$
		\State $r^{\text{mean}}_k \leftarrow \frac{1}{N_a}\sum_{j=1}^{N_a}r(s_k,a^i_k)$
		\State $r^{\text{var}}_k \leftarrow \frac{1}{N_a}\sum_{j=1}^{N_a}(r(s_k,a^i_k)-r^{\text{mean}}_k)$
		\State \Return $b_{j,1} r^{\text{mean}}_k + b_{j,2} r^{\text{var}}_k$
		\EndFunction
		
		\For{epoch $e=1$ to $N_e$}
		\State Collect the experience tuple $\langle s_k,a_k,r_{k},s_{k+1}\rangle$:
		\For{$k=1$ to $N_{train}$}
		\State $s_k \leftarrow c_k$
		\State $a_k \leftarrow \pi(s_k)$
		\State $r_k$ is obtained by Eq. (\ref{eq:crps})
		\State $s_{k+1} \leftarrow c_{k+1}$
		\EndFor
		\State Compute the discounted rewards:
		\State $Q(s_k,a_k) \leftarrow r(s_k,a_k) + \sum_{i=1}^{t}\gamma^{t-i}r(s_{k-i},a_{k-i})$
		\State Estimate the advantage:
		\State $A^{\pi_{{k}}}_{s, a} \leftarrow Q(s_k,a_k) -V^{\pi_{{k}}}(s_k)$
		\State Update the policy by minimizing the loss function:
		\State $\mathcal{L}^\text{CLIP}=\min (\frac{\pi_k(a \mid s)}{\pi_{k-1}(a \mid s)} A^{\pi_{{k}}}_{s, a},\text{clip}(\frac{\pi_k(a \mid s)}{\pi_{k-1}(a \mid s)},1-\epsilon,1+\epsilon) A^{\pi_{{k}}}_{s, a})$
		\State The details are given in the Appendix D.
		\State Update the policy by minimizing the loss function:
		\State $\mathcal{L}^\text{Q}=\mathbb{E}_{s,a \sim \pi_k}[(\gamma V^{\pi_{{k}}}(s_{k+1}) + r(s_k, a_k) -V^{\pi_{{k}}}(s_k))^2]$
		\State Using PSO to obtain the adaptive exploration.
		\EndFor
		\State \Return The well trained actor $\pi$
	\end{algorithmic}	
\end{breakablealgorithm}

\section{Comparative Metrics in Experiments}
\setcounter{equation}{0}
\renewcommand\theequation{C.\arabic{equation}}
To compare performances of the algorithms mentioned in this paper, the competitive metrics are needed to be used. The comprehensive evaluation of probabilistic prediction should consider its accuracy, variation, and reliability. Here, we adopt three metrics, i.e., CRPS, Winkler \cite{sun_using_2020} and Pinball \cite{wang_probabilistic_2019} to evaluate the three aspects, respectively. The definition of CRPS is given in Eq. (\ref{eq:crps}). As a famous metric, the Winkler is used to evaluate the variation of probabilistic distribution, and it is expressed as follows.
\begin{equation}
	\text{Winkler} = \begin{cases}
		2(\text{L}_p(\hat{y_{t}}) - y_{t})/\alpha+\delta,& \quad y_{t}<\text{L}_p(\hat{y_{t}}) \\
		2(y_{t} - \text{U}_p(\hat{y_{t}}))/\alpha+\delta,& \quad y_{t}>\text{U}_p(\hat{y_{t}}) \\
		\delta, & \quad \text{otherwise}
	\end{cases}
\end{equation}
where $\hat{y_{t}}$ and $y_t$ denote the predicted distribution and the real charging power at time $t$. $\text{L}_{p}(\hat{y_{t}})$ and $\text{U}_{p}(\hat{y_{t}})$ indicate the lower and upper quantiles at probability $p$. $\alpha$ and $\delta$ are two parameters of this metric, which are set as 0.1 and 1, respectively. Normally, the lower value of Winkler means the better variation.

{{In addition, considering the reliability of the predicted probabilistic distribution, Pinball is introduced as a metric, and its mathematical formulation is expressed as follows.
\begin{equation}
	\text{Pinball}_p = \begin{cases}
		(y_{t} - \hat{y}_{p,t})p,& \quad \hat{y}_{p,t}<y_{t}\\
		(\hat{y}_{p,t} - y_{t})(1-p),& \quad \hat{y}_{p,t}\geq y_{t}
	\end{cases}
\end{equation}
where $p$ stands for the probability of $\hat{y}_{p,t}>y_{t}$ and $\hat{y}_{p,t}$ indicates the predicted value at time $t$. The average Pinball value $\text{Pinball} = \sum_{p}\text{Pinball}_p$ of the probabilistic prediction is used to represent its performance. Note that the lower Pinball stands for the better performance.}}

\section{The Derivatives of Learnable Parameters in LSTM-AePPO}
\setcounter{equation}{0}
\renewcommand\theequation{D.\arabic{equation}}
In this section, we provide the derivatives of the learnable parameters $W_f, W_i, W_c, W_o, b_f, b_i, b_{c}, b_o$ and $\theta$ of LSTM-AePPO. 
Based on the Eqs. (\ref{eq:lstm_start})$\sim$(\ref{eq:lstm_end}), the LSTM takes $X_t$ as its input and output $y_{t+1}$ as its forecast. Its loss function is denoted as $\mathcal{L}$. The $W_f$ is taken as an example to describe its derivative\footnote{During the intermediate training of LSTM, the $\frac{\partial L^t}{\partial y^t}$ may be equal to 0 because the loss function has not calculated at time $t$. In this case, to avoid the gradient disappearance that caused by the zero of derivative, we use $\left( \frac{\partial \mathcal{L}^t}{\partial y^t} + \frac{\partial \mathcal{L}^{t+1}}{\partial y^t} \right)$ instead of $\frac{\partial \mathcal{L}^t}{\partial y^t}$.}.
\begin{equation}
\begin{split}
\frac{d \mathcal{L}}{dW_f} &= \left( \frac{\partial \mathcal{L}^t}{\partial y^t} + \frac{\partial \mathcal{L}^{t+1}}{\partial y^t} \right)\frac{\partial y^t}{\partial W_f^t}\\
&= \left(\left( \frac{\partial \mathcal{L}^t}{\partial y^t} + \frac{\partial \mathcal{L}^{t+1}}{\partial y^t} \right)\frac{\partial y^t}{\partial W_f^t}\right)+ \frac{\partial c^{t+1}}{\partial W_f^t}\\
&= \left(\left(\left( \frac{\partial \mathcal{L}^t}{\partial y^t} + \frac{\partial \mathcal{L}^{t+1}}{\partial y^t} \right)\frac{\partial y^t}{\partial c^t}\right)+ \frac{\partial c^{t+1}}{\partial c^t}\right)\frac{\partial c^{t}}{\partial W_f^t}\\
& = \left(\left(\left( \frac{\partial \mathcal{L}^t}{\partial y^t} + \frac{\partial \mathcal{L}^{t+1}}{\partial y^t} \right)\frac{\partial y^t}{\partial c^t}\right)+ \frac{\partial c^{t+1}}{\partial c^t}\right)\frac{\partial c^{t}}{\partial f_t}\frac{\partial f_t}{\partial W_f}
\end{split}
\end{equation}
where ${\partial \mathcal{L}^{t+1}}/{\partial y^t}$ is the derivative of error at $t+1$ with respect to $y^t$. In this paper, the loss function $\mathcal{L}$ is mean square error, which could be formulated by
\begin{equation}
\mathcal{L}^t = \frac{1}{2}(y^t - \hat{y}^t)^2.
\end{equation}

Then, the other derivatives are given in the following explicit formulas.
\begin{equation}
\frac{\partial \mathcal{L}^t}{\partial y^t} = y^t-\hat{y}^t
\end{equation}
\begin{equation}
\frac{\partial y^t}{\partial c^t} = o_t(1-tath(c_t)^2)
\end{equation}
\begin{equation}
\frac{\partial c^{t+1}}{\partial c^t} = f_{t+1}
\end{equation}
\begin{equation}
\frac{\partial c^{t}}{\partial f_t} = c^{t-1}f_t(1-f_t)
\end{equation}
\begin{equation}
\frac{\partial f_n}{\partial W_f}=[y_{t},X_t]
\end{equation}

Similarly, the derivative of $W_i, W_c, W_o, b_f, b_i, b_{c}$ and $b_o$ could be represented by the following equations.
\begin{equation}
\frac{d \mathcal{L}}{dW_i} = \left(\left(\left( \frac{\partial \mathcal{L}^t}{\partial y^t} + \frac{\partial \mathcal{L}^{t+1}}{\partial y^t} \right)\frac{\partial y^t}{\partial c^t}\right)+ \frac{\partial c^{t+1}}{\partial c^t}\right)\frac{\partial c^{t}}{\partial i_t}\frac{\partial i_t}{\partial W_i}
\end{equation}
\begin{equation}
\frac{d \mathcal{L}}{dW_c} = \left(\left(\left( \frac{\partial \mathcal{L}^t}{\partial y^t} + \frac{\partial \mathcal{L}^{t+1}}{\partial y^t} \right)\frac{\partial y^t}{\partial c^t}\right)+ \frac{\partial c^{t+1}}{\partial c^t}\right)\frac{\partial c^{t}}{\partial \tilde{c}_t}\frac{\partial \tilde{c}_t}{\partial W_c}
\end{equation}
\begin{equation}
\frac{d \mathcal{L}}{dW_o} = \left( \frac{\partial \mathcal{L}^t}{\partial y^t} + \frac{\partial \mathcal{L}^{t+1}}{\partial y^t} \right)\frac{\partial y^t}{\partial o_t}\frac{\partial o_t}{\partial W_o}	
\end{equation}		
\begin{equation}
\frac{d \mathcal{L}}{db_f} = \left(\left(\left( \frac{\partial \mathcal{L}^t}{\partial y^t} + \frac{\partial \mathcal{L}^{t+1}}{\partial y^t} \right)\frac{\partial y^t}{\partial c^t}\right)+ \frac{\partial c^{t+1}}{\partial c^t}\right)\frac{\partial c^{t}}{\partial f_t}\frac{\partial f_t}{\partial b_f}
\end{equation}
\begin{equation}
\frac{d \mathcal{L}}{db_i} = \left(\left(\left( \frac{\partial \mathcal{L}^t}{\partial y^t} + \frac{\partial \mathcal{L}^{t+1}}{\partial y^t} \right)\frac{\partial y^t}{\partial c^t}\right)+ \frac{\partial c^{t+1}}{\partial c^t}\right)\frac{\partial c^{t}}{\partial i_t}\frac{\partial i_t}{\partial b_i}
\end{equation}
\begin{equation}
\frac{d \mathcal{L}}{db_c} = \left(\left(\left( \frac{\partial \mathcal{L}^t}{\partial y^t} + \frac{\partial \mathcal{L}^{t+1}}{\partial y^t} \right)\frac{\partial y^t}{\partial c^t}\right)+ \frac{\partial c^{t+1}}{\partial c^t}\right)\frac{\partial c^{t}}{\partial \tilde{c}_t}\frac{\partial \tilde{c}_t}{\partial b_c}
\end{equation}
\begin{equation}
\frac{d \mathcal{L}}{db_o} = \left( \frac{\partial \mathcal{L}^t}{\partial y^t} + \frac{\partial \mathcal{L}^{t+1}}{\partial y^t} \right)\frac{\partial y^t}{\partial o_t}\frac{\partial o_t}{\partial b_o}
\end{equation}
where 
\begin{equation}
\frac{\partial c^{t}}{\partial i_t} = \tilde{c}^{t}i_t(1-i_t)
\end{equation}
\begin{equation}
\frac{\partial c^{t}}{\partial \tilde{c}_t} = i_t
\end{equation}
\begin{equation}\frac{\partial y^t}{\partial o_t} = tath(c_t)\end{equation}
\begin{equation}
\frac{\partial i_t}{\partial W_i}=\frac{\partial \tilde{c}_t}{\partial W_c}=\frac{\partial o_t}{\partial W_o}=[y_{t},X_t]
\end{equation}
\begin{equation}
\frac{\partial f_t}{\partial b_f}=\frac{\partial i_t}{\partial b_i}=\frac{\partial \tilde{c}_t}{\partial b_c}=\frac{\partial o_t}{\partial b_o} = 1	
\end{equation}

Then, the learnable parameters $\theta$ of AePPO is provided.
To simplify the notation, we denote ${\pi_t(a \mid s)}/{\pi_{t-1}(a \mid s)}$ as the probability ratio by $r(a|s)$.
Then, the Eq. (\ref{eq:actor_loss}) could be rewritten as
\begin{equation}
\begin{split}
&\mathcal{L}^{\text{CLIP}}=\\
&\mathbb{E}_{s,a \sim \pi_t}\bigg[\min (r(a|s) A^{\pi_{{t}}}_{s, a},\text{clip}(r(a|s),1-\epsilon,1+\epsilon) A^{\pi_{{t}}}_{s, a})\bigg]
\end{split}
\end{equation}

The derivative of $\theta$ could be derived from the above formula.
\begin{equation}
\frac{d\mathcal{L}^{\text{CLIP}}}{d\theta}=\mathbb{E}_{s,a \sim \pi_k}\left[\frac{\partial \log \pi^\theta_t(a|s)}{\partial \theta}r(a|s)A^{\pi_{{t}}}_{s, a}\right]
\end{equation}
where $\pi^\theta_t$ indicates the \emph{actor} at iteration $t$ with parameter $\theta$. Note that the term $r(a|s)A^{\pi_{{t}}}_{s, a}$ in the above formula is clipped into the range $[1-\epsilon,1+\epsilon]$ \cite{schulman_proximal_2017}. The $\partial \log \pi_\theta(a|s) / \partial \theta$ is formulated as follows.
\begin{equation}
\begin{split}
&\frac{\partial \log \pi_\theta(a|s)}{\partial \theta} = \\
&\frac{1}{(a^{\text{AePPO}}_{\text{var}})^2}\left[(a - a^{\theta}_\text{mean}) \left(\frac{(a - a^{\theta}_{\text{mean}})^2 - (a^{\text{AePPO}}_{\text{var}})^2}{a^{\text{AePPO}}_{\text{var}}} \right)\right]
\end{split}
\end{equation}
where $a^{\text{AePPO}}_{\text{var}}$ is obtained by the adaptive exploration mechanism rather than AePPO during the training, which is different with traditional PPO. Besides, $a^{\theta}_{\text{mean}}$ is obtained by AePPO with learnable parameters $\theta$.

{{With the formulation of these derivative, the learnable parameters could be update through the following equation with learning rate $\eta$.}}
\begin{equation}
P^t = P^t + \Delta P^t
\end{equation}
where $P^t$ indicates the learnable parameters of LSTM-AePPO and the $\Delta P^t$ is represented by
\begin{equation}
\Delta P^t = \Delta P^{t-1} + \eta \frac{\partial \mathcal{L}}{\partial P^t}
\end{equation}
where ${\partial \mathcal{L}}/{\partial P^t}$ denotes the derivative of $P^t$. 
\end{appendices}

\vspace{-1.2cm}
\scriptsize
\begin{IEEEbiography}[{\includegraphics[width=1in,height=1.25in,clip,keepaspectratio]{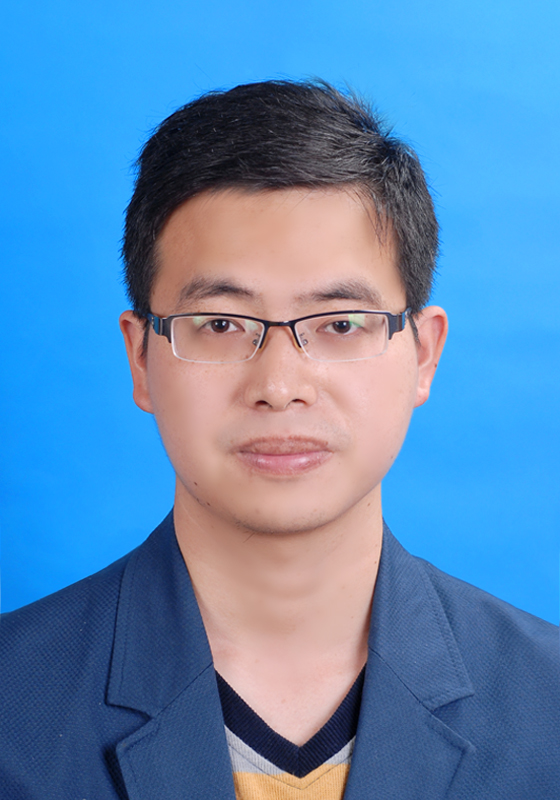}}]{Yuanzheng Li} received the M.S. degree and Ph.D. degree in Electrical Engineering from Huazhong University of Science and Technology (HUST), Wuhan, China, and South China University of Technology (SCUT), Guangzhou, China, in 2011 and 2015, respectively. Dr Li is currently an Associate Professor in HUST. He has published several peer-reviewed papers in international journals. His current research interests include electric vehicle, deep learning, reinforcement learning, smart grid, optimal power system/microgrid scheduling and decision making, stochastic optimization considering large-scale integration of renewable energy into the power system and multi-objective optimization.
\end{IEEEbiography}

\begin{IEEEbiography}[{\includegraphics[width=1in,height=1.25in,clip,keepaspectratio]{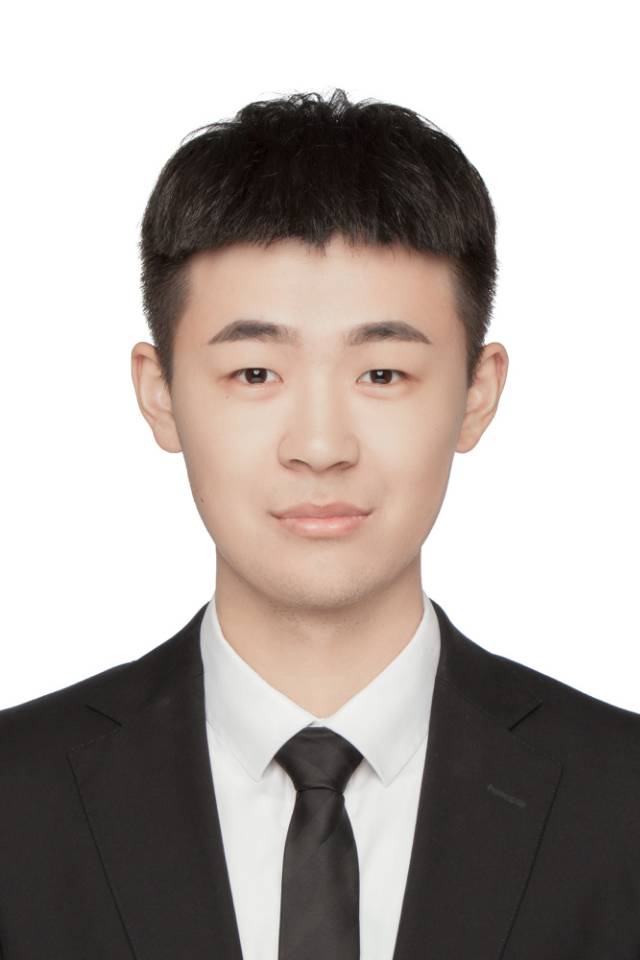}}]{Shangyang He}
received the B.S. degree in Automation from the Wuhan University of Technology in 2020. He is currently working toward the M.S. degree in Huazhong University of Science and Technology (HUST), Wuhan, China. His research interests include electric vehicle charging power forecast, deep learning, deep reinforcement learning and graph neural network.
\end{IEEEbiography}
\begin{IEEEbiography}[{\includegraphics[width=1in,height=1.25in,clip,keepaspectratio]{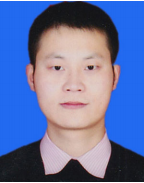}}]{Yang Li} received the Ph.D. degree in electrical engineering from North China Electric Power
University, Beijing, China, in 2014.
He is an Associate Professor with the School
of Electrical Engineering, Northeast Electric
Power University, Jilin, China. He is currently
also a China Scholarship Council-funded Postdoc with Argonne National Laboratory, Lemont,
IL, USA. His research interests include power
system stability and control, integrated energy
systems, renewable energy integration, and smart grids.
\end{IEEEbiography}

\begin{IEEEbiography}[{\includegraphics[width=1in,height=1.25in,clip,keepaspectratio]{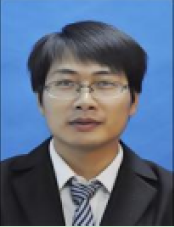}}]{Leijiao Ge} received the bachelor degree in electrical engineering and its automation from Beihua University, Jilin, China, in 2006, and the master degree in electrical engineering from Hebei University of Technology, Tianjin, China, in 2009, and the Ph.D. degree in Electrical Engineering from Tianjin University, Tianjin, China, in 2016. He is currently an associate professor in the School of Electrical and Information Engineering at Tianjin University. His main research interests include situational awareness of smart distribution network, cloud computing and big data.
\end{IEEEbiography}

\begin{IEEEbiography}[{\includegraphics[width=1in,height=1.25in,clip,keepaspectratio]{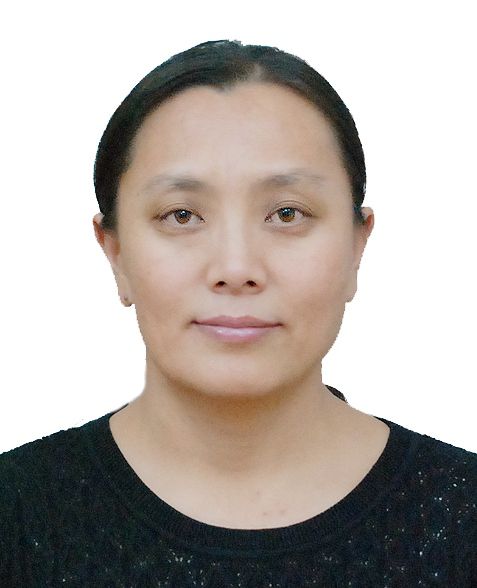}}]{Suhua Lou} received her B.E., M.Sc. and Ph.D. degrees in electrical engineering from Huazhong University of Science and Technology (HUST), Wuhan, China, in 1996, 2001 and 2005, respectively. She is currently working as a Professor in HUST. Her research interests include power system planning,energy economics and renewable energy generation. 
\end{IEEEbiography}

\begin{IEEEbiography}[{\includegraphics[width=1in,height=1.25in,clip,keepaspectratio]{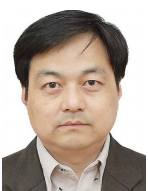}}]{Zhigang Zeng} (IEEE Fellow) received the Ph.D.
degree in systems analysis and integration from Huazhong University of Science and Technology,Wuhan, China, in 2003.
He is currently a Professor with the School of Automation and the Key Laboratory of Image Processing and Intelligent Control of the Education Ministry of China, Huazhong University of Science and Technology. He has published more
than 100 international journal articles. His current research interests include theory of functional differential equations and differential equations with discontinuous right-hand sides and their applications to dynamics of neural networks, memristive systems, and control systems. Dr. Zeng was an Associate Editor of the IEEE TRANSACTIONS ON NEURAL NETWORKS AND LEARNING SYSTEMS from 2010 to 2011. He has been an Associate Editor of the IEEE TRANSACTIONS ON CYBERNETICS since 2014 and the IEEE TRANSACTIONS ON FUZZY SYSTEMS since 2016, and a member of the Editorial Board of Neural Networks since 2012, Cognitive Computation since 2010, and Applied Soft Computing since 2013.
\end{IEEEbiography}

\end{document}